\documentclass[11pt,twoside,a4paper]{article}

\usepackage[utf8]{inputenc}
\usepackage[T1]{fontenc}
\usepackage{microtype}
\usepackage{amsmath,amsfonts,amssymb,mathtools}
\usepackage{graphicx}
\usepackage{bm}
\usepackage[margin=0.9in]{geometry}
\usepackage{siunitx}
\usepackage{setspace}
\usepackage{xcolor}

\usepackage[square,numbers,sort&compress]{natbib}
\newcommand{\tavg}[1]{\left\langle #1 \right\rangle}
\newcommand{\davg}[1]{\left\langle #1 \right\rangle_{\bm{J}}}

\usepackage{fancyhdr}
\usepackage[colorlinks,
    citecolor=blue,
    filecolor=black,
    urlcolor=black,
    linkcolor=blue,
    linktoc=all]{hyperref}
\usepackage[font=footnotesize]{caption}

\usepackage{authblk}

\begin{document}

\title{\bf \Large Structure of activity in multiregion recurrent neural networks}
\author{David G. Clark\thanks{\href{mailto:dgc2138@cumc.columbia.edu}{dgc2138@cumc.columbia.edu}}\:}
\author{Manuel Beiran\thanks{\href{mailto:mb4878@columbia.edu}{mb4878@columbia.edu}}}

\affil{Zuckerman Institute, Columbia University, New York, NY, USA}

\date{\today}

\maketitle

\spacing{1.1}

\begin{abstract}
Neural circuits comprise multiple interconnected regions, each with complex dynamics. The interplay between local and global activity is thought to underlie computational flexibility, yet the structure of multiregion neural activity and its origins in synaptic connectivity remain poorly understood. We investigate recurrent neural networks with multiple regions, each containing neurons with random and structured connections. Inspired by experimental evidence of communication subspaces, we use low-rank connectivity between regions to enable selective activity routing. These networks exhibit high-dimensional fluctuations within regions and low-dimensional signal transmission between them. Using dynamical mean-field theory, with cross-region currents as order parameters, we show that regions act as both generators and transmitters of activity---roles that are often in tension. Taming within-region activity can be crucial for effective signal routing. Unlike previous models that suppressed neural activity to control signal flow, our model achieves routing by exciting different high-dimensional activity patterns through connectivity structure and nonlinear dynamics. Our analysis offers insights into multiregion neural data and trained neural networks.
\end{abstract}

\newpage
\tableofcontents
\newpage 

\section{Introduction}

A striking example of convergent evolution in nervous systems is the emergence of well-defined anatomical regions that interact with one another \citep{felleman1991distributed, Ito2014, Randlett2015, wang2020allen}. Recent advances in neural-recording technologies have enabled simultaneous monitoring of thousands of neurons across multiple brain areas \textit{in vivo} \citep{Jun2017, Machado2022, manley2024simultaneous, chen2024brain}. These studies reveal that neurons exhibit varying degrees of regional specialization in their activities \citep{markov2011weight, Ecker2014, Lin2015, wang2020allen}. This regional specialization, balanced with cross-region interactions, is believed to underlie the flexible, adaptive capabilities of neural circuits \citep{perich2020inferring, okazawa2023, fang2023predictive}. Modern neural datasets thus reveal an intricate interplay between region-specific and broadly distributed signals.

These datasets raise fundamental questions about the origins and functions of multiregion neural activity \citep{musall2019single, Steinmetz2019, BrainLaboratory2023, schaffer2023spatial}. To address them, researchers have trained multiregion recurrent neural network models, either to perform cognitive tasks \citep{pinto2019task, michaels2020goal, chen2021modularity, barbosa2023early} or to generate recorded neural data \citep{Andalman2019, nair2023approximate}. These models have shed light on directed multiregion interactions involved in sensorimotor processing, context modulation, and changes in behavioral states \citep{perich2020rethinking}.

However, in both real neural circuits and their artificial counterparts, the nature of multiregion interactions remains largely mysterious. In particular, we lack understanding of the connectivity supporting modular computations and the mechanisms of flexible signal routing. The coexistence and interaction of region-specific and network-wide dynamics are also unclear.

To address these challenges, we analyze a recurrent network model with multiple regions. Each region has a combination of random and low-rank connectivity, generating both high-dimensional fluctuations and specific low-dimensional patterns \citep{mastrogiuseppe2018linking, Pereira-Obilinovic2023}. We connect regions using low-rank connectivity, enabling selective routing of low-dimensional signals between regions.

Due to its nonlinear dynamics and multiregion connectivity structure, this model produces an extremely rich and broad array of dynamic states depending on the connectivity. We develop an analytical theory of this multiregion activity structure by deriving and solving dynamical mean-field theory (DMFT) equations for the network in the limit where each region has infinitely many neurons for any finite number of regions. Given the complexity of the resulting DMFT equations, we solve them in stages of increasing complexity: first considering symmetric effective interactions leading to fixed-point solutions in the low-dimensional dynamics, then progressing to include disorder. Finally, we examine general effective interactions with the potential for limit-cycle solutions, requiring numerical solution.

Our analysis reveals two key ideas, each supported by various specific results:

\textbf{Key idea 1:} Regions serve dual roles as generators and transmitters of activity, with an inherent tension between these functions. When the intrinsic dynamics within a region become too strong or complex, the region's ability to transmit signals is compromised. Our analysis characterizes this conflict and demonstrates how taming within-region dynamics is crucial for network-level communication.

\textbf{Key idea 2:} Signal routing throughout the network is achieved by shifting which subspaces of high-dimensional activity space are excited or unexcited through the interplay of connectivity statistics and nonlinear recurrent dynamics. The subset of subspaces that are excited depends on the geometric arrangement of low-rank connectivity patterns and the strength of disordered connectivity. Our approach complements earlier models of gating and routing in neural circuits, which emphasized single-neuron biophysical mechanisms such as neuromodulation, inhibition, or gain modulation \cite{abbott2006switches}, by developing a geometric, population-level view of information flow.

Overall, our work provides a theoretical framework for understanding the interplay between regional specialization and multiregion interactions in neural circuits, offering insights into the mechanisms underlying flexible signal routing and modular computations in the brain.

\section{Multiregion Network Model}

Here, we present the multiregion network model, first describing its dynamics and then its connectivity. 

\subsection{Dynamics}

We study rate-based (non-spiking) recurrent neural networks comprising $R$ regions, each containing $N$ neurons. We consider a finite number of regions $R$ and take the limit $N \rightarrow \infty$, corresponding to a small or moderate number of regions, each with a large number of neurons. The preactivations of the neurons, analogous to membrane potentials, are denoted by $x_i^\mu(t)$, where $\mu \in \{1,\ldots,R\}$ specifies the region and $i \in \{1,\ldots,N\}$ specifies the within-region neuron. The activations, analogous to firing rates, are given by $\phi_i^\mu(t) = \phi(x_i^\mu(t))$, where $\phi(x) = \text{erf}(\sqrt{\pi}x/2)$ is a pointwise nonlinearity that is linear for small $|x|$ and saturates at $\pm 1$ for large $|x|$. Neurons interact through a synaptic coupling matrix $J^{\mu\nu}_{ij}$ according to:
\begin{equation}
    \frac{dx_i^\mu(t)}{dt} = -x_i^\mu(t) + \sum_{\nu=1}^R \sum_{j=1}^N J_{ij}^{\mu\nu} \phi_j^\nu(t).
    \label{eq:network-dynamics}
\end{equation}

\subsection{Connectivity}
\begin{figure}[t!]
    \centering
    \includegraphics[width=1.85in]{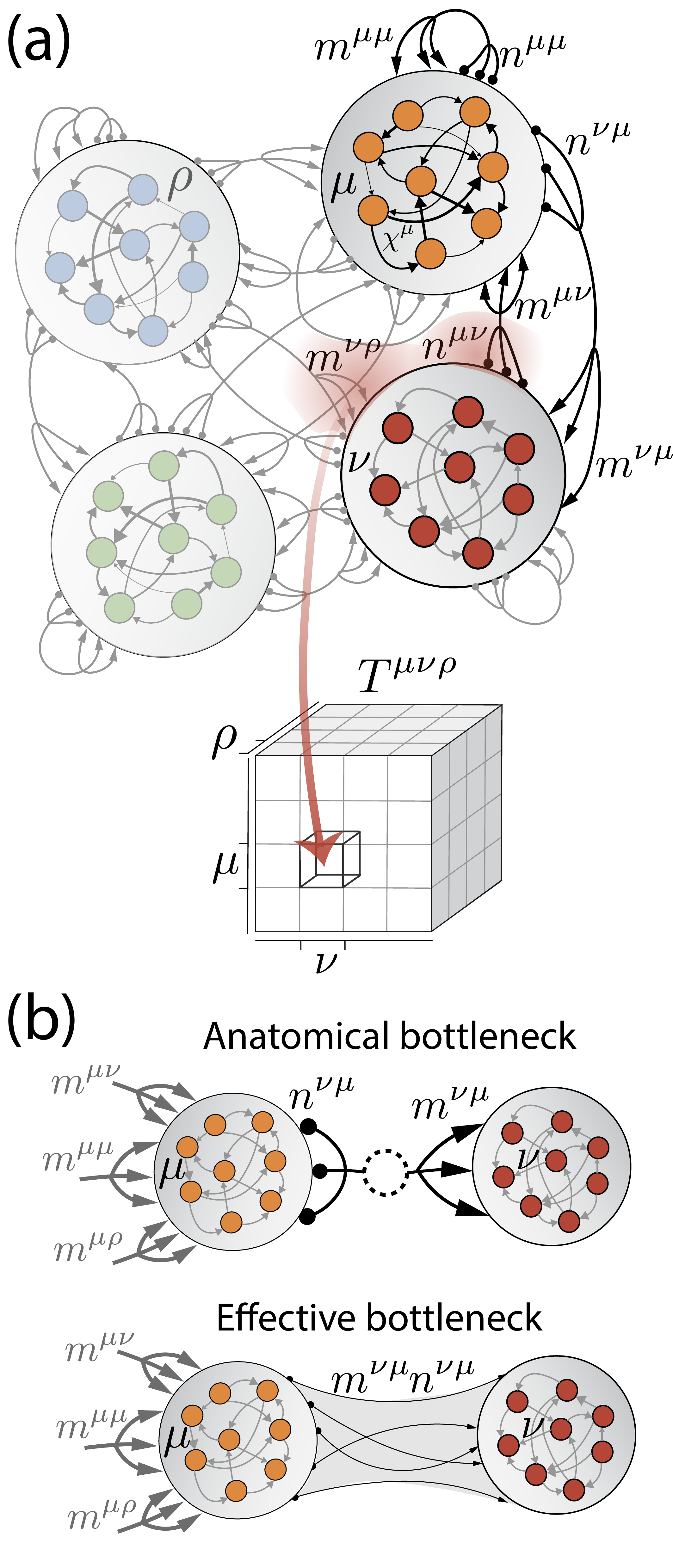}
    \caption{(a) Top: Schematic of the synaptic connectivity model. Different regions, each with ``random plus rank-one'' connectivity, are linked via rank-one matrices representing communication subspaces. In this network of $R=4$ regions, we highlight the rank-one and disordered couplings in region $\mu$, as well as the structured couplings to and from region $\nu$. Rank-one connections are defined through the outer product of vectors $\bm{m}^{\mu \nu}$ and $\bm{n}^{\mu\nu}$. Bottom: Tensor $T^{\mu\nu\rho}$, which encodes the geometric arrangement of the connectivity patterns and determines the dynamics of region-to-region currents in the mean-field picture. (b) Anatomical bottleneck or effective bottleneck implementing a rank-one connectivity matrix between regions $\nu$ and $\mu$. The dashed circle represents a linear neuron with fast timescale.}
    \label{fig:schematic}
\end{figure}

The connections within each region $\mu$ are dense and consist of the sum of random disordered couplings, $\chi^\mu_{ij}$, and a rank-one matrix, as investigated by Mastrogiuseppe and Ostojic \cite{mastrogiuseppe2018linking}. This rank-one matrix is defined as the outer product of vectors $\bm{m}^{\mu\mu}$ and $\bm{n}^{\mu\mu}$. Connections between pairs of regions, such as from region $\nu$ to $\mu$, are represented by additional rank-one matrices formed by outer products of vectors $\bm{m}^{\mu\nu}$ and $\bm{n}^{\mu\nu}$ (Fig.~\ref{fig:schematic}a). The synaptic coupling matrix is thus expressed as:
\begin{equation}\label{eq:connJ}
    J^{\mu\nu}_{ij} = \delta^{\mu\nu}\chi^\mu_{ij} + \frac{1}{N} m^{\mu\nu}_i n^{\mu\nu}_j.
\end{equation}
Each element of $\chi^\mu_{ij}$ is sampled independently from a zero-mean Gaussian with variance $(g^\mu)^2/N$. This $1/\sqrt{N}$ scaling of the disordered couplings ensures that the eigenspectrum of $\chi^{\mu}_{ij}$ remains independent of network size for large $N$.

For tractability, we assume that the components of the vectors $\bm{m}^{\mu\nu}$ and $\bm{n}^{\mu\nu}$ are zero-mean random variables drawn from a multivariate Gaussian. Specifically, for each neuron in the network, such as for neuron $i$ in region $\nu$, there are $2R$ jointly sampled components: $\{n_i^{\mu\nu}\}_{\mu=1}^R \cup \{m_i^{\nu\rho}\}_{\rho=1}^R$. To define the second-order statistics of these components, we introduce the tensors:
\begin{subequations}
\begin{align}
    T^{\mu\nu\rho} &= \davg{n_i^{\mu \nu} m_i^{\nu \rho}}, \\
    U^{\mu \nu \rho} &= \davg{m_i^{\mu \nu} m_i^{\mu \rho}}.
\end{align}
\end{subequations}
Our analysis will demonstrate that specifying the remaining second-order statistics, $\davg{ n_i^{\mu \nu} n_i^{\rho \nu}}$, is not necessary to study the dynamics in the limit $N\to\infty$. However, to sample the vectors defining the low-rank part of the couplings, we must specify $\davg{ n_i^{\mu \nu} n_i^{\rho \nu} }$. We set this proportional to $\delta^{\mu \rho}$ with a scale factor large enough to ensure that the overall covariance matrix of vector components is positive-definite. As $N\to\infty$, these tensors can equivalently be expressed by the normalized overlaps or inner products:
\begin{subequations}
\begin{align}
    T^{\mu\nu\rho} &= \frac{1}{N} \sum_{i=1}^N {n_i^{\mu \nu} m_i^{\nu \rho}}, \label{eq:T_def} \\
    U^{\mu \nu \rho} &= \frac{1}{N} \sum_{i=1}^N {m_i^{\mu \nu} m_i^{\mu \rho}} \label{eq:U_def}.
\end{align}
\end{subequations}
Thus, $T^{\mu\nu\rho}$ and $U^{\mu\nu\rho}$ encode the geometric arrangement of connectivity patterns (Fig.~\ref{fig:schematic}a, bottom), providing a concise representation of the network's structure. When showing simulation results, we will consider only large networks where the particular realization of connectivity is not significant, and the system behavior is controlled by $g^\mu$, $T^{\mu\nu\rho}$, and $U^{\mu\nu\rho}$.

Table~\ref{table:1} summarizes the variables and notation used throughout this article.

\section{Biological Motivations and Assumptions}

In constructing this model, we aimed to incorporate sufficient biological detail to capture nontrivial phenomena while maintaining analytical tractability. In this section, we elucidate the biological foundations of our model, outlining its underlying assumptions and limitations, first addressing the dynamics and then the connectivity.

\subsection{Dynamics: Motivation and Assumptions}

The complexity in our network model's dynamics, compared to linear networks that can simply be diagonalized, stems from the nonlinear activations of individual neurons. This nonlinearity is inspired by the transformation of input currents into spike trains by real neurons. While our model captures this crucial aspect, it does not account for other features of cortical circuits, such as distinct excitatory and inhibitory populations (i.e., Dale's law), sparse connectivity, and nonnegative firing rates.

This level of abstraction mirrors that used in the seminal work of Sompolinsky et al. \cite{sompolinsky1988chaos}, which described chaotic activity arising from strong random connectivity. Indeed, our multiregion model reduces to $R$ independent samples of this model when the structured low-rank couplings are set to zero. In this special case, each disconnected region transitions from quiescence to high-dimensional chaos at a critical coupling variance, defined by $g^\mu = 1$.

Our use of this level of abstraction is supported by recent studies demonstrating that network models incorporating the biological features we omitted (i.e., nonnegative rates or spikes, sparse connections, and excitatory-inhibitory populations) can exhibit equivalent dynamical regimes. This equivalence has been observed both for disordered couplings, where the same transition to chaos occurs \cite{kadmon2015transition, mastrogiuseppe2017intrinsically}, and for low-rank couplings \cite{herbert2022impact, shao2023relating}.

\subsection{Connectivity: Motivation and Assumptions}

We use rank-one matrices to model structured connectivity both within and between regions, based on separate experimental observations for each type of connectivity.

Within-region recordings show that neural activity during tasks often lies on a low-dimensional manifold \cite{gallego2017neural, mastrogiuseppe2018linking}. Rank-one connectivity can generate arbitrary one-dimensional dynamics \cite{beiran2021shaping}, serving as a starting point for modeling structured low-dimensional activity. Many standard neural-network models, including Hopfield networks \cite{hopfield1982neural}, ring attractors \cite{ben1995theory}, and autoencoders \cite{deneve2017brain}, use low-rank connectivity. Furthermore, our model combines rank-one and disordered within-region connectivity. As shown by Mastrogiuseppe and Ostojic \cite{mastrogiuseppe2018linking}, such networks can produce chaotic activity, fixed points, or both, depending on the relative strengths of rank-one vs. disordered connectivity.

Cross-region rank-one connections are based on observed \textit{communication subspaces} between cortical areas. In particular, Semedo et al. \cite{semedo2019cortical} found that only a low-dimensional subspace of V1 activity, distinct from the subspace capturing most V1 variance, correlates with activity in V2. Similar communication-subspace structure has been identified in visual processing \cite{semedo2022feedforward}, motor control \cite{perich2020motor, kondapavulur2022transition}, attention \cite{srinath2021attention}, audition \cite{barbosa2023early}, and brain-wide activity \cite{macdowell2023multiplexed}. Low-rank cross-region connectivity offers a simple explanation for these subspaces, but of course is not the only explanation. Alternative hypotheses, such as global fluctuations or shared input, were considered less likely based on anatomy, spatial selectivity, and persistence under anesthesia by the authors of the original study (in visual cortex). Here, we adopt low-rank connectivity for its simplicity, data compatibility, and, as we discuss in the next section, functional utility.

Biologically, low-rank cross-region connectivity, which acts as a type of bottleneck, can be implemented either anatomically or effectively (Fig.~\ref{fig:schematic}b; \cite{sussillo2009generating, mastrogiuseppe2018linking}). An anatomical bottleneck would involve a set of intermediary neurons between two areas (Fig.~\ref{fig:schematic}b, top). These neurons, assumed to be linear with fast time constants, would read out activity from the source region and broadcast it to the target region \cite{logiaco2021thalamic}. This framework also accommodates thalamocortical loops as anatomical bottlenecks between cortical regions (this complements existing models where thalamic nuclei create loops within a cortical area; such loops can be selectively modulated via basal-ganglia inhibition, controlling inter-region communication). Alternatively, an effective bottleneck would arise from direct, monosynaptic connections between source and target regions with a low-rank structure (Fig.~\ref{fig:schematic}b, bottom). A simple example of this occurs when all connections from a source to a target region have the same strength and sign, corresponding to a rank-one matrix that is sensitive only to the mean activity of the source region.

Under the interpretation of an effective bottleneck, the rank-one constraint results in a synaptic coupling from a neuron in region $\nu$ to a neuron in region $\mu$ that is proportional to the product of two scalar variables: $n^{\mu \nu}_i$ and $m^{\mu \nu}_j$. These variables are associated with the emitter and receiver populations, respectively. Such couplings, expressed as products of pre- and postsynaptic terms, arise naturally in neuroscience as a consequence of Hebbian plasticity.

Finally, while we use rank-one matrices, a more realistic model might involve higher-rank matrices, or matrices with smoothly decaying singular values. We find that even rank-one matrices induce rich multiregion activity structure, providing an adequate starting point. 

\subsection{Functional Significance of Low-Rank Cross-Region Connectivity}

A rank-one connectivity matrix implements an activity-dependent bottleneck: the transmission of activity from source region $\nu$ to target region $\mu$ depends on the alignment of activity in $\nu$ with the row space of the connecting low-rank matrix. This row space, given by the span of $\bm{n}^{\mu\nu}$, represents the communication subspace in our model. The bottleneck then projects this filtered activity into target region $\mu$ through the column space of the matrix, given by the span of $\bm{m}^{\mu\nu}$.

This connectivity structure allows selective communication between regions, controlled by the geometry encoded in $T^{\mu\nu\rho}$. To illustrate this mechanism, consider an activity pattern $\phi^\nu_i$ in region $\nu$. The activity communicated to region $\mu$ is proportional to the projection $N^{-1}\sum_{i=1}^N n^{\mu\nu}_i \phi^\nu_i$. For a generic pattern $\phi^\nu_i$ (e.g., induced by the disordered connectivity $\chi^\nu_{ij}$), this projection is of order $1/\sqrt{N}$, vanishing as $N \rightarrow \infty$. However, if $\phi^\nu_i$ has a component aligned with $\bm{n}^{\mu\nu}$, this projection remains of order unity.

For such alignment to occur, there must exist a region $\rho$ such that $\bm{m}^{\nu\rho}$, which delivers input to region $\nu$, has a component along $\bm{n}^{\mu\nu}$. This component is precisely $T^{\mu\nu\rho}$. Consequently, high-dimensional chaotic activity cannot propagate between regions as $N \rightarrow \infty$, ensuring that only structured, low-dimensional signals are transmitted.

\begin{table}[t!]
\centering
\footnotesize
\begin{tabular}{|p{2cm}|p{13cm}|}
\hline
\multicolumn{2}{|c|}{Network variables} \\
\hline
$x_i^\mu(t)$ & Preactivation (``membrane potential'') of neuron $i$ in region $\mu$ at time $t$ (Eq.~\ref{eq:network-dynamics}) \\
$\phi_i^\mu(t)$ & Activation (``firing rate'') of neuron $i$ in region $\mu$ at time $t$ (Eq.~\ref{eq:network-dynamics}) \\
\hline
\multicolumn{2}{|c|}{Network parameters} \\
\hline
$N$ & Number of neurons in each region \\
$R$ & Number of regions \\
$J_{ij}^{\mu\nu}$ & Synaptic coupling from neuron $j$ in region $\nu$ to neuron $i$ in region $\mu$ (Eq.~\ref{eq:connJ}) \\
$\chi_{ij}^{\mu}$ & Random component of within-region synaptic couplings in region $\mu$ (Eq.~\ref{eq:connJ})\\
$g^{\mu}$ & Standard deviation (times $\sqrt{N}$) of random couplings in region $\mu$ \\
$\bm{m}^{\mu \nu}$ & Vector with components $m_i^{\mu \nu}$; defines structured input pattern from region $\nu$ to neurons in region $\mu$ (Eq.~\ref{eq:connJ}) \\
$\bm{n}^{\mu \nu}$ & Vector with components $n_i^{\mu \nu}$; defines structured readout pattern from neurons in region $\nu$ to region $\mu$ (Eq.~\ref{eq:connJ}) \\
\hline
\multicolumn{2}{|c|}{DMFT variables} \\
\hline
$\Delta^{\mu}(t,t')$ & Correlation function of preactivations in region $\mu$ (Eq.~\ref{eq:Delta_def}) \\
$C^{\mu}(t,t')$ & Correlation function of activations in region $\mu$ (Eq.~\ref{eq:C_def}) \\
$S^{\mu \nu}(t)$ & Current from region $\nu$ to region $\mu$ at time $t$ (Eq.~\ref{eq:S_def}) \\
$H^{\mu\nu}(t)$ & Drive to $S^{\mu \nu}(t)$ in the mean-field dynamics of the currents (Eq.~\ref{eq:current-dynamics}) \\
$\psi^{\mu}(t)$ & Neuronal gain in region $\mu$ at time $t$ (Eq.~\ref{eq:psi-integral}) \\
$A^{\mu}$ & Sum of squared currents from all regions into region $\mu$ \\
$S_0^{\mu \nu}$ & Fixed-point value of inter-region current from region $\nu$ to region $\mu$ \\
$\sigma^{\mu \nu}(t)$ & Perturbation to the inter-region current from region $\nu$ to region $\mu$ \\
$\hat{\Delta}^{\mu}(\tau)$ & Normalized stationary correlation function (Eq.~\ref{eq:Delta_hat_def}) \\
\hline
\multicolumn{2}{|c|}{DMFT parameters} \\
\hline
$T^{\mu \nu \rho}$ & Normalized overlap between readout and input patterns, representing effective interaction from region $\rho$ to region $\mu$ through region $\nu$ (Eq.~\ref{eq:T_def}) \\
$\hat{T}^{\mu \nu, \rho \sigma}$ & Matrix form of $T^{\mu\nu\rho}$ (Eq.~\ref{eq:matrix_form}) \\
$U^{\mu \nu \rho}$ & Overlap between input vectors in region $\mu$ originating from regions $\nu$ and $\rho$ (Eq.~\ref{eq:U_def}) \\
$c^{\mu \nu}$ & Symmetric parameterization of $T^{\mu\nu\rho}$ (Eq.~\ref{eq:sym-form}) \\
$u^{\mu}$ & Rank-one contribution to ``rank-one plus diagonal'' parameterization of $c^{\mu\nu}$ (Eq.~\ref{eq:c_rank_one}) \\
$h^{\mu}$ & Diagonal contribution to ``rank-one plus diagonal'' parameterization of $c^{\mu\nu}$ (Eq.~\ref{eq:c_rank_one}) \\
$a^{\mu}$ & Strength of direct self-interaction (Eq.~\ref{eq:ab_def}) \\
$b^{\mu}$ & Strength of indirect self-interaction (Eq.~\ref{eq:ab_def}) \\
\hline
\end{tabular}
\caption{Summary of notation.}\label{table:1}
\end{table}

\section{Dynamical Mean-Field Theory (DMFT)}

Mean-field theory is an analytical approach that describes large systems using a small set of summary statistics called \textit{order parameters}. This method provides an exact description as $N \rightarrow \infty$ and a good approximation for large, finite $N$. Dynamical mean-field theory (DMFT) extends this concept by introducing time-dependent order parameters to capture the temporal evolution of activity \cite{sompolinsky1988chaos, hansel1993solvable}. We now present the order parameters in the DMFT description of our multiregion network model and the equations governing their dynamics.

\subsection{Order Parameters}

Our multiregion model exhibits two types of dynamics: high-dimensional chaotic fluctuations from i.i.d. connectivity, and low-dimensional excitation within or between regions due to low-rank connectivity. These dynamics are described by distinct sets of order parameters.

High-dimensional fluctuations are characterized by correlation functions, which capture the temporal structure of chaotic fluctuations. For each region $\mu$, we define correlation functions for the (pre-)activations:
\begin{subequations}
\begin{align}
    \Delta^\mu(t,t') &= \frac{1}{N} \sum_{i=1}^N x^\mu_i(t) x^\mu_i(t'), \label{eq:Delta_def} \\
    C^\mu(t,t') &= \frac{1}{N} \sum_{i=1}^N \phi^\mu_i(t) \phi^\mu_i(t') \label{eq:C_def}.
\end{align}
\label{eq:region-specific-twopt-fns}\end{subequations}

Low-dimensional signal transmission within and between regions is described by \textit{currents}, following the terminology of Perich et al. \cite{perich2020inferring}. These currents are consolidated in the matrix $S^{\mu\nu}(t)$, defined by:
\begin{equation}
    \frac{d S^{\mu\nu}(t)}{dt} = -S^{\mu\nu}(t) + \frac{1}{N}\sum_{i=1}^N n^{\mu\nu}_i \phi_i^\nu(t). \label{eq:S_def}
\end{equation}
The current $S^{\mu\nu}(t)$ represents the activity in region $\nu$ that is transmitted to region $\mu$ (plus a low-pass filter).

\subsection{Routing and Non-Routing Regions}

The current matrix provides crucial information about activity flow between regions. We classify regions as routing or non-routing based on their role in signal transmission. We say that a region $\nu$ is \textbf{routing} if it transmits signals between other regions, indicated by at least one nonzero off-diagonal element in the $\nu$-th column of the current matrix, $S^{:,\nu}(t)$; and at least one nonzero off-diagonal element in the $\nu$-th row, $S^{\nu,:}(t)$. In contrast, we say that a region $\nu$ is \textbf{non-routing} if all elements of its corresponding row and column in the current matrix are zero, except possibly for the diagonal element, $S^{\nu\nu}(t)$.

As we will demonstrate through exact solutions of the DMFT equations, a region may become non-routing when its own activity is too strong, preventing signal flow. One way for this to occur is if the region's activity aligns with its internal structured connectivity, resulting in a nonzero diagonal element, $S^{\nu\nu} \neq 0$.

Experimentally, routing of this type could be detected through analyses similar to those used by Semedo et al. \cite{semedo2019cortical}. By computing the communication subspace for a source region during spontaneous activity, one could see how activity patterns line up with that subspace during a task; the overlapping activity would be the routed signal.

\subsection{Dynamical Mean-Field Equations}
In the mean-field picture, currents interact according to:
\begin{subequations}
\begin{align}
    \frac{d S^{\mu\nu}(t)}{dt} &= - S^{\mu\nu}(t) + H^{\mu \nu}(t), \text{ where} \\
    H^{\mu \nu}(t) &= \psi^\nu(t) \sum_{\rho=1}^R {T}^{\mu \nu \rho} S^{ \nu\rho}(t),
\end{align}\label{eq:current-dynamics}\end{subequations}
where $\psi^\nu(t) = \psi(\Delta^\nu(t,t))$ is the average gain of neurons in region $\nu$. The function $\psi(\Delta)$ performs a Gaussian average:
\begin{equation}
    \psi(\Delta) = \tavg{\phi'(x)}_x, 
    \label{eq:psi-integral}
\end{equation}
where $x \sim \mathcal{N}(0, \Delta)$.
Thus, while standard neural networks have a vector dynamics shaped by a matrix, in our framework, region-to-region interactions, defined by the current order parameters, have a matrix dynamics shaped by a third-order tensor.
Meanwhile, $\Delta^\mu(t,t')$ satisfies:
\begin{equation}
    \left(1+\frac{d}{dt}\right)\left(1 + \frac{d}{dt'} \right)\Delta^{\mu}(t, t') = (g^{\mu})^2 C^\mu(t, t') 
    + \sum_{\nu,\rho=1}^R U^{\mu\nu\rho}H^{\mu \nu}(t) H^{\mu \rho}(t'),
    \label{eq:two-pt-fn-dynamics}
\end{equation}
These equations are closed by expressing $C^\mu(t,t')$ in terms of $\Delta^\mu(t,t')$ via $C^\mu(t,t') = C(\Delta(t,t'), \Delta(t,t), \Delta(t',t'))$, where $C(\Delta_{12}, \Delta_{11}, \Delta_{22})$ propagates preactivation correlations to activation correlations:
\begin{equation}
     C(\Delta_{12}, \Delta_{11}, \Delta_{22}) = \tavg{\phi(x_1)\phi(x_2)}_{x_1,x_2},
    \label{eq:C-integral}
\end{equation}
where $(x_1,x_2) \sim \mathcal{N}\left(\bm{0}, \bm{\Delta}\right)$. $\psi(\Delta)$ and $C(\Delta_{12}, \Delta_{11}, \Delta_{22})$ can be evaluated analytically (SI Appendix).

Thus, the DMFT provides a set of deterministic, causal dynamic equations for the region-specific two-point functions and currents. While their derivation is relatively straightforward, solving them analytically is challenging due to their nonlinear and time-dependent structure, as well as the tensorial form of the interactions. In the next section, we show that by assuming certain symmetry properties of $T^{\mu\nu\rho}$, we can, remarkably, derive a rich and instructive class of time-independent and time-dependent solutions.

For the remainder of the paper, we assume $U^{\mu\nu\rho} = \delta^{\nu\rho}$ for all $\mu$, focusing on the role of $T^{\mu\nu\rho}$. Geometrically, this means that inputs from other regions into a target region $\mu$ are organized in orthogonal subspaces.

\section{Symmetric Effective Interactions and Fixed Points}

\begin{figure}[t!]
    \centering
    \includegraphics[width=3.25in]{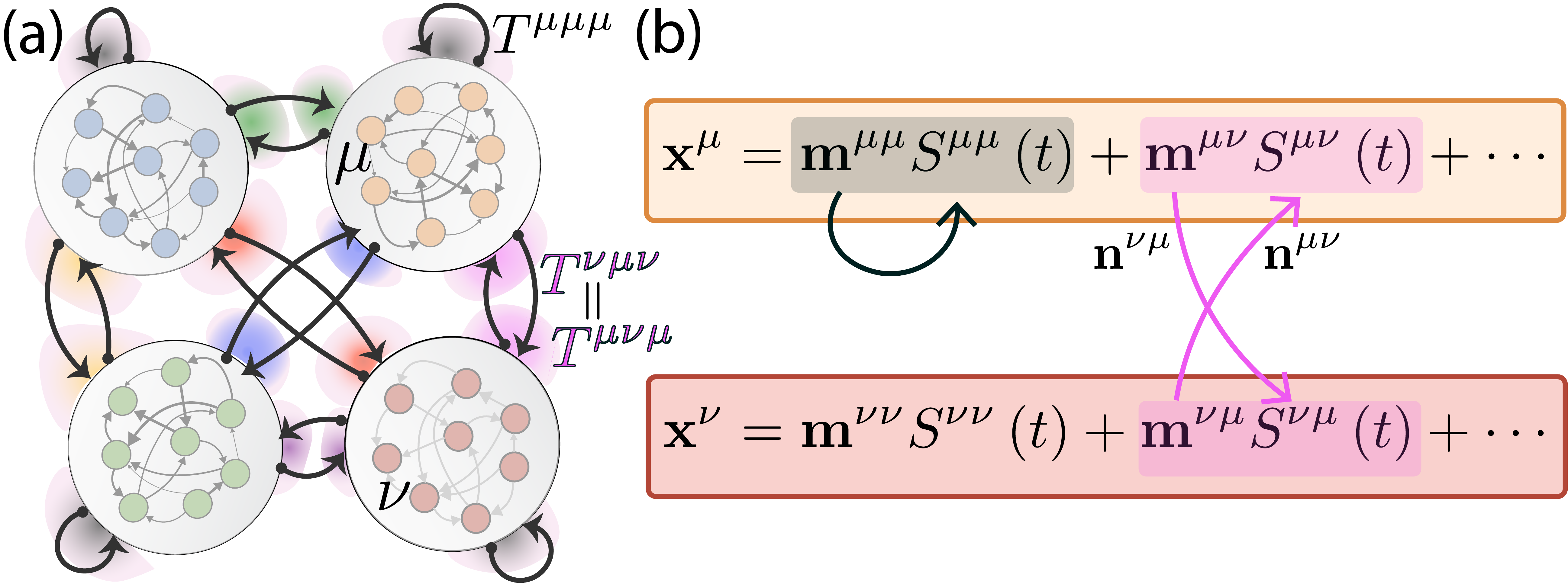}
    \caption{(a) Restriction to the effective-interaction tensor $T^{\mu\nu\rho}$ corresponding to enforcing symmetry. This constraint sets $T^{\mu\nu\rho}=\delta^{\mu\rho} c^{\mu\nu}$, where $c^{\mu\nu}$ is a symmetric matrix. nonzero overlaps between connectivity patterns are indicated by colored auras, with equal colors indicating equal overlaps. In this scenario with $R = 4$ regions, the connectivity has 10 independent parameters: 4 for direct and 6 for indirect effective self-interactions. (b) Illustration of subspace-based routing in the case of symmetric effective interactions. When the activity subspace defined by the span of ${m}_i^{\mu \mu}$ in region $\mu$ is excited, bidirectional communication between regions $\mu$ and $\nu$ is suppressed, and vice versa, due to the nonlinear dynamics of the network.}
    \label{fig:sym-connectivity}
\end{figure}

We now set out to derive exact solutions to the DMFT equations. In general, to simplify the analysis of many-body interactions, a natural choice is to assume \textit{symmetry}. In standard neural networks, symmetric interactions ensure that the system converges to fixed points, precluding limit cycles and chaos. However, enforcing symmetry in the DMFT system is challenging because the effective interactions among the currents form a third-order tensor, $T^{\mu\nu\rho}$.

To clarify the structure of the interactions between currents in the DMFT, we rewrite the right-hand side of the current dynamics as
$
    \psi^\nu(t) \sum_{\rho, \sigma=1}^R \hat{T}^{\mu\nu,\rho\sigma} S^{\rho \sigma}(t),
$
where
\begin{equation}
   \hat{T}^{\mu\nu,\rho\sigma} = \delta^{\nu\rho} T^{\mu\nu\sigma} \label{eq:matrix_form}
\end{equation}
is a $R^2$-by-$R^2$ dynamics matrix governing the linearized interaction of the $R^2$ currents (its spectrum is closely related to that of $J^{\mu\nu}_{ij}$; SI Appendix). We expect $\hat{T}^{\mu\nu,\rho\sigma}$ to influence the current dynamics similarly to how the synaptic weight matrix shapes neuronal dynamics in a standard neural network. Thus, a natural choice is to impose symmetry on the matrix $\hat{T}^{\mu\nu,\rho\sigma}$, i.e., $\hat{T}^{\mu\nu,\rho\sigma} = \hat{T}^{\rho\sigma,\mu\nu}$. This reduces the number of free parameters from $\mathcal{O}(R^3)$ to $\mathcal{O}(R^2)$ by requiring
\begin{equation}
    T^{\mu\nu\rho} = \delta^{\mu\rho} c^{\mu\nu}, \text{  where  } c^{\mu\nu} = c^{\nu\mu}.
    \label{eq:sym-form}
\end{equation}
The presence of $\delta^{\mu\rho}$ in $T^{\mu\nu\rho}$ implies that each region $\mu$ interacts either directly with itself ($\mu = \nu$) or indirectly with itself through an intermediate region, $\nu$ ($\mu \neq \nu$). Moreover, the symmetry of $c_{\mu\nu}$ implies that the coupling through which region $\mu$ interacts with itself via region $\nu$ is equivalent to that for region $\nu$ interacting with itself via region $\mu$. This is illustrated in Fig.~\ref{fig:sym-connectivity}a.

To make analytical progress, we further constrain the symmetric matrix $c^{\mu\nu}$ to have a ``rank-one plus diagonal'' form, with only $\mathcal{O}(R)$ parameters,
\begin{equation}
c^{\mu\nu} = u^\mu u^\nu + \delta^{\mu\nu} h^\mu, \label{eq:c_rank_one}
\end{equation}
where $u^\mu$ and $h^\mu$ are arbitrary vectors.
This form provides a minimal setting in which one has independent control over the strength of direct versus indirect self-interactions, which are captured by the quantities
\begin{equation}
a^\mu = (u^\mu)^2 + h^\mu, \text{  and  } b^\mu = (u^\mu)^2, \label{eq:ab_def}
\end{equation}
respectively.
If $b^\mu = 0$, region $\mu$ is not connected to the rest of the network, and its dynamical repertoire is that of a rank-one network with disorder, studied in \cite{mastrogiuseppe2018linking}.

\subsection{Disorder-Free Case}

\begin{figure}[t!]
    \centering
    \includegraphics[width=3.25in]{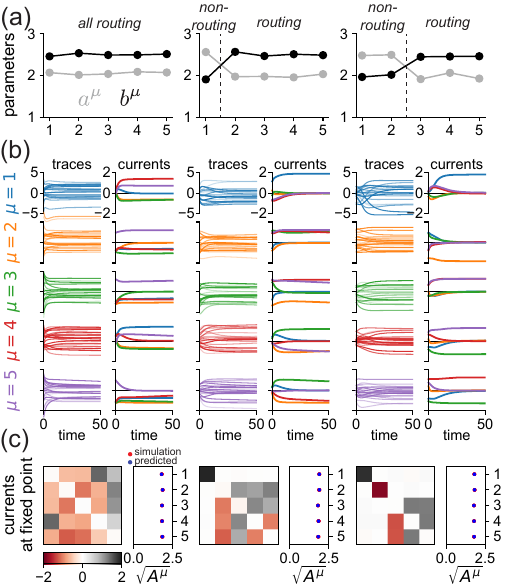}
    \caption{Structure of fixed points in networks with symmetric effective interactions. The same information for three different cases is shown on the left, center and right. (a) Values of $a^\mu$ and $b^\mu$ in the $R=5$ regions. (b) Dynamics of sampled neurons (left) and of incoming currents (right) in large simulations for each region. (c) Visualization of the steady-state current matrix $S^{\mu \nu}_0$ (left) and of the $L^2$-norms of the rows of this matrix (right). We show row-norms from the simulations (red dots) alongside analytical predictions (blue dot). In the leftmost plots, all regions are in non-routing mode. In the middle plots, region 1 is in non-routing mode and regions 2--4 are in routing mode. In the rightmost plots, regions 1 and 2 are in non-routing mode and regions 3--5 are in routing mode.}
    \label{fig:symmetric}
\end{figure}

We begin by examining the case without disorder in connectivity: $g^\mu = 0$ for all $\mu$. Symmetric interactions typically lead to fixed points, which we find to be the case here (although we were unable to derive a global Lyapunov function). For the parameterization of $T^{\mu\nu\rho}$ defined above, the fixed points $S^{\mu\nu}_0$ of the currents satisfy:
\begin{subequations}\label{eq:fp-nodisorder-cond}
\begin{align}
    S_0^{\mu\nu} &= \psi_0^\nu (u^\mu u^\nu + \delta^{\mu\nu}h^\nu) S_0^{\nu\mu},\label{eq:fp-nodisorder-cond_a} \\
    \psi^\nu_0 &= \psi(A^\nu), \text{  where  }
    A^\nu = \sum_{\rho=1}^R {(S_0^{\nu\rho})^2}.
    \end{align}\label{eq:fp-nodisorder-cond_b}\end{subequations}
Here, $A^\mu$ represents the squared $L^2$-norm of row $\mu$ of the current matrix. In the absence of disorder, $A^\mu$ is the variance of preactivations in region $\mu$. (Note that with a general form of $U^{\mu\nu\rho}$, this would become a Mahalanobis norm.) These equations yield a combinatorial family of stable and unstable fixed points, which can be categorized based on whether each region is routing or non-routing. Notably, within this family of fixed points, \textit{a region is routing if, and only if, it produces no self-exciting activity, i.e., $S^{\mu\mu} = 0$}. This directly illustrates Key Idea 1: the tension between signal generation and transmission.

For a given fixed point, let $\mathcal{S}_\text{route} \subseteq \{1,\ldots,R\}$ be the subset of regions in routing mode.
For a region $\mu \notin \mathcal{S}_\text{route}$, Eq.~\ref{eq:fp-nodisorder-cond_a} simplifies to:
\begin{subequations}
\begin{align}
    \psi_0^{\mu} &= \frac{1}{a^{\mu}}, \\
    (S_0^{\mu \mu})^2 &= A^\mu, \\
    S_0^{\mu\nu} &= S_0^{\nu\mu} = 0 \text{ for all } \nu \neq \mu.
\end{align}\label{eq:s_nonroute}\end{subequations}
On the other hand, for a region $\mu \in \mathcal{S}_\text{route}$, Eq.~\ref{eq:fp-nodisorder-cond} implies:
\begin{subequations}
\begin{align}
    \psi_0^{\mu} &= \frac{1}{b^{\mu}}, \\
    S_0^{\mu \mu} &= 0, \\
    S_0^{\mu\nu}u^\nu &= S_0^{\nu\mu}u^\mu \text{ for all } \nu \in \mathcal{S}_\text{route} \setminus \{\mu\}.
\end{align}\label{eq:s_route}\end{subequations}
Additionally, for each region $\mu \in \mathcal{S}_\text{route}$:
\begin{equation}
    A^\mu = \sum_{\nu \in \mathcal{S}_\text{route}\setminus \{\mu\}} (S_0^{\mu\nu})^2. \label{eq:group-constr}
\end{equation}
Combining these results, we have:
\begin{equation}
A^\mu = 
\begin{cases}
\psi^{-1}\left(\frac{1}{a^{\mu}}\right) & \text{for } \mu \notin \mathcal{S}_\text{route} \\
\psi^{-1}\left(\frac{1}{b^{\mu}}\right) & \text{for } \mu \in \mathcal{S}_\text{route}.
\end{cases}
\end{equation}
Here, $\psi^{-1}(1/x) = 2 (x^2-1)/\pi$ is a monotonically increasing function of $x$, so $A^\mu$ increases with $a^\mu$ or $b^\mu$. These equations determine the row norms $A^\mu$ for all $\mu$ and the pattern of \text{(non-)zero} entries in the current matrix for a given bipartition of routing and non-routing regions. For regions in routing mode, there is remaining freedom in choosing the current-matrix off-diagonal entries, resulting in a manifold of fixed points. We analyze the dimension and topology of this manifold in the SI Appendix, finding that the set of stable fixed points (see below) forms multiple disconnected continuous attractors in current space, with the number depending on the values of $A^\mu$. 

\subsection{Stability Analysis}

There are $2^R$ possible ways to assign routing and non-routing modes to regions, producing a combinatorial class of fixed points. To determine which states are stable, we perform a stability analysis, finding that region $\mu$ is in routing mode if, and only if, $a^\mu < b^\mu$. To demonstrate this, we consider a first-order perturbation $\sigma^{\mu\nu}(t)$ about a fixed point $S_0^{\mu\nu}$ and define a ``local energy'':
\begin{equation}
E[\bm{\sigma}] = \frac{1}{2} \sum_{\mu,\nu=1}^R \left(\frac{\sigma^{\mu\nu}}{u^\mu} \right)^2.
\end{equation}
We show in the SI Appendix that $\partial_t E \leq 0$ for all $\sigma^{\mu\nu}$ if and only if $S^{\mu\nu}_0$ is in a configuration claimed to be stable. Moreover, when $S^{\mu\nu}_0$ is stable, there exists a family of choices for $\sigma^{\mu\nu}$ that lead to $\partial_t E = 0$. These directions correspond to translation along a continuous attractor manifold.

In this setup, a region $\mu$ can be toggled between routing and non-routing modes by adjusting the relative magnitudes of $a^\mu$ and $b^\mu$ (Fig.~\ref{fig:symmetric}). This approach to routing contrasts with traditional methods that manipulate individual neurons or synapses through neuromodulation, inhibition, or gain modulation. In particular, the gain $\psi^\mu_0$ is nonzero in both routing and non-routing modes, unlike conventional gain-modulation methods that would be analogous to driving $\psi_0^\mu$ to zero to achieve a non-routing state. Through the interplay between connectivity geometry and nonlinear recurrent dynamics, our model aligns neural activity with subspaces that either facilitate or inhibit cross-region communication, reflecting Key Idea 2.

\subsection{Effect of Disorder}
\label{subsec:effect-of-disorder}

\begin{figure}[t!]
    \centering
    \includegraphics[width=3.25in]{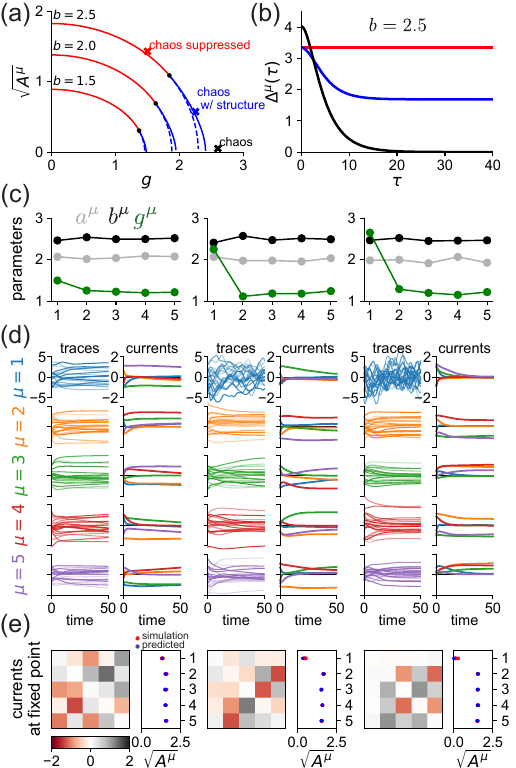}
    \caption{Structure of activity in networks with disorder and symmetric effective interactions among regions. (a) Relationship between $A^\mu$ and $g^\mu$ for various values of $b^\mu$ in the DMFT. Dashed lines indicate nonphysical solutions of the DMFT equations corresponding to unstable fixed points. (b) Solutions for the two-point function $\Delta^\mu(\tau)$ for the parameter values indicated by the markers in (a). (c--e) are the same as (a--c) in Fig.~\ref{fig:symmetric},  but with disorder, whose levels are shown in (a). All regions have $g^\mu > 1$, so regions produce high-dimensional fluctuations unless tamed by current-based activity. In the leftmost plots, chaos is suppressed in all regions, and all regions are in routing mode. In the middle plots, all regions are in routing mode, and high-dimensional fluctuations exist alongside the structured current-based activity in region 1. In the rightmost plots, region 1 is in disorder-dominated non-routing mode, and regions 2--5 are in routing mode. In chaotic regimes (middle and right columns), the inter-region currents converge to steady values despite ongoing chaotic dynamics. This convergence occurs because the readout patterns project out the chaotic fluctuations, though small $\mathcal{O}(1/\sqrt{N})$ fluctuations remain around the mean-field values.}
    \label{fig:symmetric-disorder}
\end{figure}

Maintaining the simplified parameterization of $T^{\mu\nu\rho}$, we now introduce disorder into the model by allowing nonzero values of $g^\mu$. This addition potentially leads to high-dimensional chaotic fluctuations. While these fluctuations cannot propagate through the rank-one cross-region couplings (up to small, $\mathcal{O}(1/\sqrt{N})$ fluctuations around the mean-field currents), they can disrupt low-dimensional signal transmission between regions, illustrating the tension between signal generation and transmission, Key Idea 1.

Despite the presence of disorder, the symmetric structure of the interactions ensures that the currents converge to fixed points, $S^{\mu\nu}_0$. However, the network's behavior is now controlled not just by the values of $a^\mu$ and $b^\mu$, but also by the disorder strength $g^\mu$. This richer dynamical landscape is naturally characterized by the correlation function $\Delta^{\mu}(t, t')$, which captures, for example, how quickly the network forgets its state at a given time through chaotic mixing. We focus on stationary solutions where $\Delta(t,t') = \Delta(\tau)$, with $\tau = t-t'$. Under these conditions, we can solve the DMFT equations analytically, determining $\Delta^\mu(0)$, $\Delta^\mu(\infty) = \lim{\tau \rightarrow \infty} \Delta^\mu(\tau)$, and $A^{\mu}$ (Figs.~\ref{fig:symmetric-disorder}(a) and (b); SI Appendix).

The solutions exhibit the following structure, as depicted in Figs.~\ref{fig:symmetric-disorder}(c--e). For small $g^\mu$, high-dimensional fluctuations are absent in region $\mu$, resulting in $\Delta^\mu(\tau) = \Delta^\mu(0) = \Delta^\mu(\infty)$. This constant correlation function indicates that neural activity maintains perfect memory of its state, reflecting purely structured, non-chaotic dynamics. Routing and non-routing modes behave as in the disorder-free case (Eqs.~\ref{eq:s_nonroute}--\ref{eq:group-constr}), with current stability determined by the relative magnitudes of $a^\mu$ and $b^\mu$. Here, we assume that $b^\mu > a^\mu$ so that, without disorder, all regions are in routing mode (the behavior we will describe as disorder is increased is similar for $b^\mu < a^\mu$, but with changes to self-current rather than cross-region current).

This non-chaotic regime persists even for $g^\mu > 1$, demonstrating that currents from within the region (non-routing mode) or from other regions (routing mode) can suppress chaos. However, compared to the disorder-free case, $A^\mu$ is reduced, indicating that disorder impedes currents. As $g^\mu$ increases further, a phase transition occurs. High-dimensional fluctuations begin to coexist with currents, characterized by $\Delta^\mu(\infty) < \Delta^\mu(0)$ and a decaying $\Delta^\mu(\tau)$. The decay of $\Delta^\mu(\tau)$ to a nonzero value $\Delta^\mu(\infty)$ indicates that the network partially forgets its state through chaotic mixing, while maintaining some structure through the persistent currents. In this regime, $A^\mu$ decreases even more.

At sufficiently large $g^\mu$, another phase transition takes place, leading to a ``disorder-dominated'' non-routing mode. Here, $\Delta^\mu(\tau)$ decays from $\Delta^\mu(0) > 0$ to $\Delta^\mu(\infty) = 0$, and $A^\mu = 0$. The complete decay of the correlation function indicates that the network completely forgets its state at any given time, reflecting fully chaotic dynamics with no underlying structure. The values of $\psi_0^\mu$ and $\Delta^\mu(\tau)$ are no longer influenced by $a^\mu$ and $b^\mu$. Instead, $\Delta^\mu(\tau)$ follows the solution described by Sompolinsky et al. \cite{sompolinsky1988chaos}, as if no structured connectivity were present. This disorder-dominated phase differs from the ``structure-dominated'' non-routing mode of the disorder-free case in a crucial way: signal transmission from other regions is impeded by high-dimensional fluctuations rather than structured self-exciting activity, resulting in $S^{\mu\mu}_0 = 0$.

Importantly, these disorder-induced phase transitions occur independently across regions, a consequence of the low-rank structure of cross-region connectivity preventing the propagation of high-dimensional fluctuations.

To summarize, the behavior of $\Delta^\mu(\tau)$ reveals how network activity aligns with different subspaces: when $\Delta^\mu(\tau)$ is constant, activity lies in structured subspaces defined by currents; when it decays to a nonzero value, activity combines both current-based structure and chaotic components; and when it decays to zero, activity explores all dimensions chaotically. This progression illustrates Key Idea 2: signal routing is achieved not by silencing regions, but by controlling which subspaces of activity are excited or suppressed through the interplay of connectivity and dynamics.

\section{Asymmetric Effective Interactions}
\label{sec:asymmetric-int}

We now relax all constraints on the effective interactions, including symmetry, allowing $T^{\mu\nu\rho}$ to have arbitrary elements. This can lead to a richer set of dynamic behaviors in the network. To analyze these dynamics, we focus on the spectrum of $\hat{T}^{\mu\nu,\rho\sigma}$, the matrix representation of $T^{\mu\nu\rho}$.

The leading eigenvalue of $\hat{T}^{\mu\nu,\rho\sigma}$ strongly influences the network's behavior. When this eigenvalue is real, the currents typically converge to fixed points. In contrast, a complex-conjugate pair of leading eigenvalues, especially with a large imaginary part, often results in limit cycles in the currents. We have not observed chaotic attractors in the currents.

To characterize the interplay between current dynamics, within-region high-dimensional fluctuations, and the leading eigenvalue of $\hat{T}^{\mu\nu,\rho\sigma}$, we conducted a comprehensive analysis. We focused on networks with $R=2$ regions, setting disorder levels $g^1 = g^2 = 1.5$. For each complex number $\lambda$ on a grid in the upper half-plane, we generated 50 random effective-interaction tensors $T^{\mu\nu\rho}$ whose associated matrix $\hat{T}^{\mu\nu,\rho\sigma}$ had $\lambda$ as its leading eigenvalue. For each tensor, we numerically solved the DMFT equations to obtain the two-point functions $\Delta^{\mu}(t,t')$ and currents $S^{\mu\nu}(t)$. We then analyzed the normalized two-point function:
\begin{equation}
    \hat{\Delta}^\mu(\tau) = \frac{\Delta^\mu(t,t+\tau)}{\sqrt{ \Delta^\mu(t) \Delta^\mu(t+\tau) }}, \:\:\: t \gg 1, \label{eq:Delta_hat_def}
\end{equation}
where $t$ is large enough to disregard transients. The behavior of $\hat{\Delta}^\mu(\tau)$ indicates the presence and nature of high-dimensional fluctuations in region $\mu$. In particular, similar to the interpretation of $\Delta^\mu(\tau)$ in the previous section, when $\hat{\Delta}^\mu(\tau)$ decays to a nonzero value, region $\mu$ displays chaotic fluctuations with underlying structure due to currents providing order-one mean activity. This structure can also be seen in the currents themselves. Conversely, $\hat{\Delta}^\mu(\tau)$ decaying to zero indicates that there are only chaotic fluctuations in region $\mu$.

\begin{figure}[t!]
    \centering
    \includegraphics[width=6.2in]{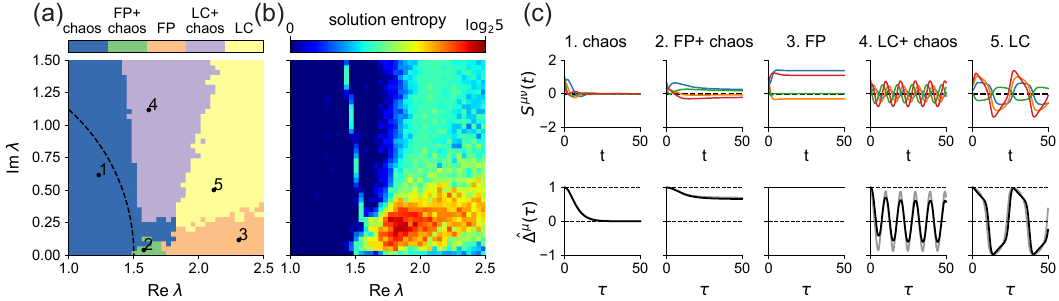}
    \caption{Dynamic behaviors in networks with asymmetric effective interactions ($R=2$ regions). (a) Most common dynamic behavior across 50 realizations of $T^{\mu\nu\rho}$, as a function of the leading eigenvalue $\lambda$ of $\hat{T}^{\mu\nu,\rho\sigma}$. (b) Entropy of the distribution over dynamic behaviors at each $\lambda$. (c) Example time series of currents $S^{\mu \nu}$ (top) and two-point functions $\hat{\Delta}^\mu(\tau)$ (bottom) for each dynamic behavior. In the top row, colors represent different currents; in the bottom row, black and gray lines represent the two regions.}
    \label{fig:limit-cycles}
\end{figure}

Figure \ref{fig:limit-cycles} summarizes our findings. As the real part of $\lambda$ increases with a small imaginary part, we observe a progression from pure chaos, to fixed points coexisting with chaos, to pure fixed points (Fig.~\ref{fig:limit-cycles}a,c). Strikingly, when the imaginary part of $\lambda$ is larger, we see a parallel series of transitions: from chaos, to limit cycles coexisting with chaos, to pure limit cycles. The coexistence of limit cycles with high-dimensional fluctuations is particularly intriguing, as it demonstrates that reliable, time-dependent routing can occur beneath apparently noisy activity.

The dashed circle in Fig.~\ref{fig:limit-cycles}a indicates the support of the bulk spectrum of $J^{\mu\nu}_{ij}$. For nontrivial current dynamics to emerge, the leading eigenvalue of $\hat{T}^{\mu\nu,\rho\sigma}$ must lie outside this circle. This illustrates how high-dimensional fluctuations within regions (the bulk) can impede structured cross-region communication (the outlier), highlighting the tension between signal generation and transmission (Key Idea 1).

To assess the predictive power of the leading eigenvalue, we computed the entropy of the empirical distribution over the five possible dynamic states at each $\lambda$ (Fig.~\ref{fig:limit-cycles}b). For large imaginary parts of $\lambda$, we observe a reliable transition from chaos to limit cycles coexisting with high-dimensional fluctuations as the real part increases, with a critical value near $\text{Re}\lambda = 1.5$. In regions where pure fixed points or limit cycles dominate, the behavior becomes more variable, especially where different states intermingle.

\begin{figure}[t!]
    \centering
    \includegraphics[width=3.25in]{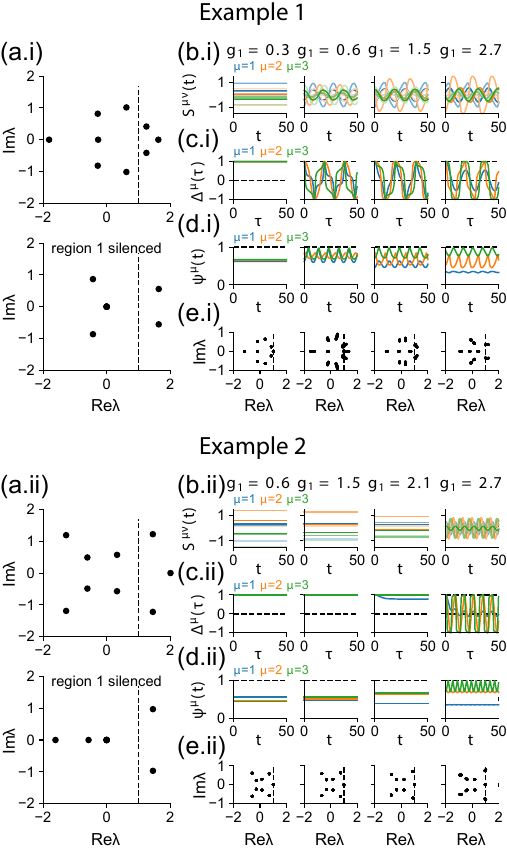}
    \caption{Modulating multiregion dynamics through disorder in a 3-region network. Two examples (1 and 2) show how introducing disorder in region 1 switches current dynamics from fixed points to limit cycles. (a) Spectra of $\hat{T}^{\mu\nu,\rho\sigma}$ before (top) and after (bottom) silencing region 1. The resulting switch from real to complex-conjugate pair of the leading eigenvalue suggests that introducing disorder in region 1 will generate limit cycles. (b) Time evolution of currents $S^{\mu\nu}$, with colors indicating the target region $\mu$. (c) Normalized two-point functions $\hat{\Delta}^\mu(\tau)$ for increasing disorder $g^1$ in region 1. (d) Time-dependent gains $\psi^\mu(t)$. (e) Time-evolving spectra of $\psi^\nu(t) \hat{T}^{\mu\nu,\rho\sigma}$, showing how the eigenvalue distribution changes throughout the limit cycle.}
    \label{fig:asymm}
\end{figure}

We next explored how modulating disorder can shape multiregion dynamics and signal routing. Figure \ref{fig:asymm} shows two cases with fixed ${T}^{\mu\nu\rho}$ in networks of $R = 3$ regions. In both cases, introducing disorder in region 1 switched the current dynamics from fixed points to limit cycles. Importantly, this transition did not occur by silencing region 1; instead, the gains of all regions remained of order unity throughout the transition (Fig.~\ref{fig:asymm}c). This supports Key Idea 2, demonstrating that signal routing is achieved by shaping the alignment of neural activity with particular subspaces, rather than through traditional gain modulation methods. 

To further understand time-dependent signal routing, we analyzed the spectrum of $\psi^\nu(t) \hat{T}^{\mu\nu,\rho\sigma}$ across time (Fig.~\ref{fig:asymm}d). During limit cycles, the leading eigenvalues hover around unity, indicating that current dynamics are regulated through sequential subspace activation and subtle gain adjustments.

These findings demonstrate that in both fixed-point and dynamic attractor scenarios, adjusting effective interactions or disorder levels can shift signal routing through the network. This routing occurs not by silencing entire regions, but by altering which subspaces are active, leading to phase transitions in current dynamics while maintaining nonzero gains. This mechanism aligns with both Key Ideas 1 and 2, highlighting the tension between signal generation and transmission and emphasizing the role of subspace activation in controlling signal flow.

\section{Input-Driven Switches}

Our model shows that a region's ability to transmit signals depends on the balance between its within-region activity and cross-region communication, as described in Key Idea 1. While this balance can be modified by adjusting synaptic couplings, as demonstrated in the previous sections, external inputs offer an alternative method for controlling routing that is more amenable to experimental probing \cite{barbosa2023early}. 

We extended the DMFT to incorporate inputs, introducing new effective interactions that capture overlaps between recurrent connectivity and input vectors (SI Appendix). To illustrate this, we examined a simple example with 5 regions. Initially, region 1 exhibits strong self-exciting activity and does not route signals. When we add input to region 1 that other regions can read out and feed back, it transitions to a state where region 1 communicates with the network and its self-exciting activity is suppressed. This input-driven switch mirrors the connectivity-based switches studied earlier and exemplifies one of many possible scenarios for input-based activity modulation.

The specific effects of inputs depend on the multiregion connectivity geometry encoded in $T^{\mu\nu\rho}$. Experimentally, inputs could be provided to a region using techniques like optogenetics. Given knowledge of cross-region subspace geometry, one could predict resulting network-level activity changes. This geometry could be estimated using methods similar to those developed by Semedo et al. \cite{semedo2019cortical}.

\section{Discussion}
\label{sec:discussion}

In this work, we focused on rank-one communication subspaces with jointly Gaussian loadings. This connectivity provides a starting point for studying more complicated forms of communication between areas. For example, we can extend our rank-one connectivity model to rank-$K$ subspaces, facilitating richer, higher-dimensional communication. Maintaining the ranks of these subspaces as intensive prevents high-dimensional chaotic fluctuations from propagating between regions, preserving the modularity of the disorder-based gating mechanism. While increasing the rank increases the number of dynamic variables in the mean-field picture (namely, by a factor of $K$), the Gaussian distribution determining the loadings restricts the complexity of their effective interactions. An alternative is to use a mixture-of-Gaussians distribution with $C$ components, allowing for more complex interactions, such as chaotic dynamics among the currents \cite{beiran2021shaping, dubreuil2022role}. Together, these extensions expand the effective-interaction tensor by three indices, detailed in a tensor diagram in the SI Appendix. Finally, an important future direction will be to incorporate biological constraints, such as excitatory and inhibitory neurons and nonnegative firing rates. The work of \cite{kadmon2015transition} is a promising starting point.

How might the connectivity geometry defining $T^{\mu\nu\rho}$ be established? We propose that this structure could emerge through the pressures of a learning process. Consider a region $\mu$ that needs to perform a computation based on a one-dimensional signal from region $\nu$. In this case, establishing a rank-one cross-region coupling matrix $\bm{m}^{\mu\nu} (\bm{n}^{\mu\nu})^T$, which could occur through Hebbian plasticity, is sufficient. The preactivations in $\nu$ lie within the subspace $S^\nu = \text{span}\{\bm{m}^{\nu \rho}\}_{\rho=1}^R$. For $\mu$ to use a signal from $\nu$, the row space spanned by $\bm{n}^{\mu\nu}$ must then overlap with $S^\nu$. This overlap implies that $T^{\mu\nu\rho} = N^{-1} (\bm{n}^{\mu\nu})^T \bm{m}^{\nu\rho} \neq 0$ for at least one $\rho$. This simplified picture of learning neglects the fact that regions are connected in loops. Future research is required to explore how regions learn tasks in a recurrently connected network, addressing the ``multiregion credit assignment'' problem.

The question ``What defines a brain region?'' is, at its essence, about how within-region connectivity differs from cross-region connectivity. Previous work, such as that by Aljadeff et al. \cite{aljadeff2015transition}, studied networks with disordered couplings both within and between regions, but found that chaotic activity is globally distributed, undermining the notion of distinct regions. In contrast, our model, which uses low-rank cross-region connectivity, leads to rich functional consequences and modular activity states, making it a more interesting candidate framework for regional organization.

The symmetric connectivity geometry we studied, characterized by $c^{\mu\nu}$, has not yet been observed in functional communication-subspace analyses or current connectomics data. However, as larger-scale mammalian connectomes become available in the coming years, it would be valuable to compute observables like $T^{\mu\nu\rho}$. Given its interesting functional consequences, our symmetry-constrained version would be a natural structure to look for, analogous to how researchers have examined correlations between reciprocal synapses in existing datasets.

A notable aspect of our model and theoretical approach is its alignment with existing methods for neural-data analysis. Specifically, the technique developed by Perich et al. \cite{perich2020inferring} for analyzing multiregion neural recordings involves training a recurrent network to mimic the data, then decomposing the activity in terms of cross-region currents. Intriguingly, our model's low-dimensional mean-field dynamics offer a closed description in terms of these currents, rather than relying solely on single-region quantities such as two-point functions. This alignment strongly supports the use of current-based analyses in neural data interpretation.

Furthermore, our model could be adapted to fit multiregion neural data using approaches akin to those of Valente et al. \cite{valente2022extracting}. Subsequently reducing the model to the mean-field description we derived could provide insights into the dynamics of the fitted model. This positions our work as a bridge connecting practical recurrent network-based data analysis methods to a deeper analytical understanding of network dynamics.

Another data-driven application of our framework lies in analyzing connectome data \cite{abbott2020mind}. Large-scale reconstructions of neurons and their connections are now available for flies \cite{zheng2018complete, scheffer2020connectome}, parts of the mammalian cortex \cite{winnubst2019reconstruction}, and other organisms \cite{hildebrand2017whole}. For connectome datasets where regions are identified, the cross-region connectivity could be approximated as having a low-rank structure, allowing for a reduction using our mean-field framework. This enables a comparison of predicted neuronal dynamics with recorded activity.

In scenarios where regions are not already defined, our framework suggests solving the ``inverse problem'': determining a partitioning of neurons into regions such that the cross-region connectivity is well approximated by low-rank matrices. Developing a specialized clustering algorithm for this purpose and applying it to connectome data, such as from the fly, would be interesting. Even in cases where anatomical knowledge suggests certain region definitions, identifying ``unsupervised regions'' based on the assumption of low-rank cross-region interactions could offer an interesting new functional perspective on regional delineation.

\section*{Acknowledgments}

We are extremely grateful to L.F. Abbott for his advice on this work. We thank Albert J. Wakhloo for comments on the manuscript, as well as Rainer Engelken, Haim Sompolinsky, Ashok Litwin-Kumar, and members of the Litwin-Kumar and Xiao-Jing Wang groups for helpful discussions. D.G.C. was supported by the Kavli Foundation. M.B. was supported by NIH award R01EB029858. The authors were additionally supported by the Gatsby Charitable Foundation GAT3708.

\newpage

\appendix 

\section{Appendix}

\subsection{Spectral analysis}
\label{sec:spectral-analysis}

We describe the spectrum of $J^{\mu\nu}_{ij}$, which controls the local dynamics about the trivial fixed point, $x^{\mu}_{i} = 0$. Rather than as a fourth-order tensor, $J^{\mu\nu}_{ij}$ can be regarded as an $RN$-by-$RN$ matrix with respect to the ``superindices'' $(\mu,i)$ and $(\nu,j)$.
This $RN$-by-$RN$ matrix has spectral bulk from the i.i.d. matrices $\chi_{ij}^\mu$, whose density in the complex plane, for $N \rightarrow \infty$, is a superposition of $R$ uniform disks of radii $g^\mu$.

\begin{figure}[h]
    \centering
    \includegraphics[width=3.5in]{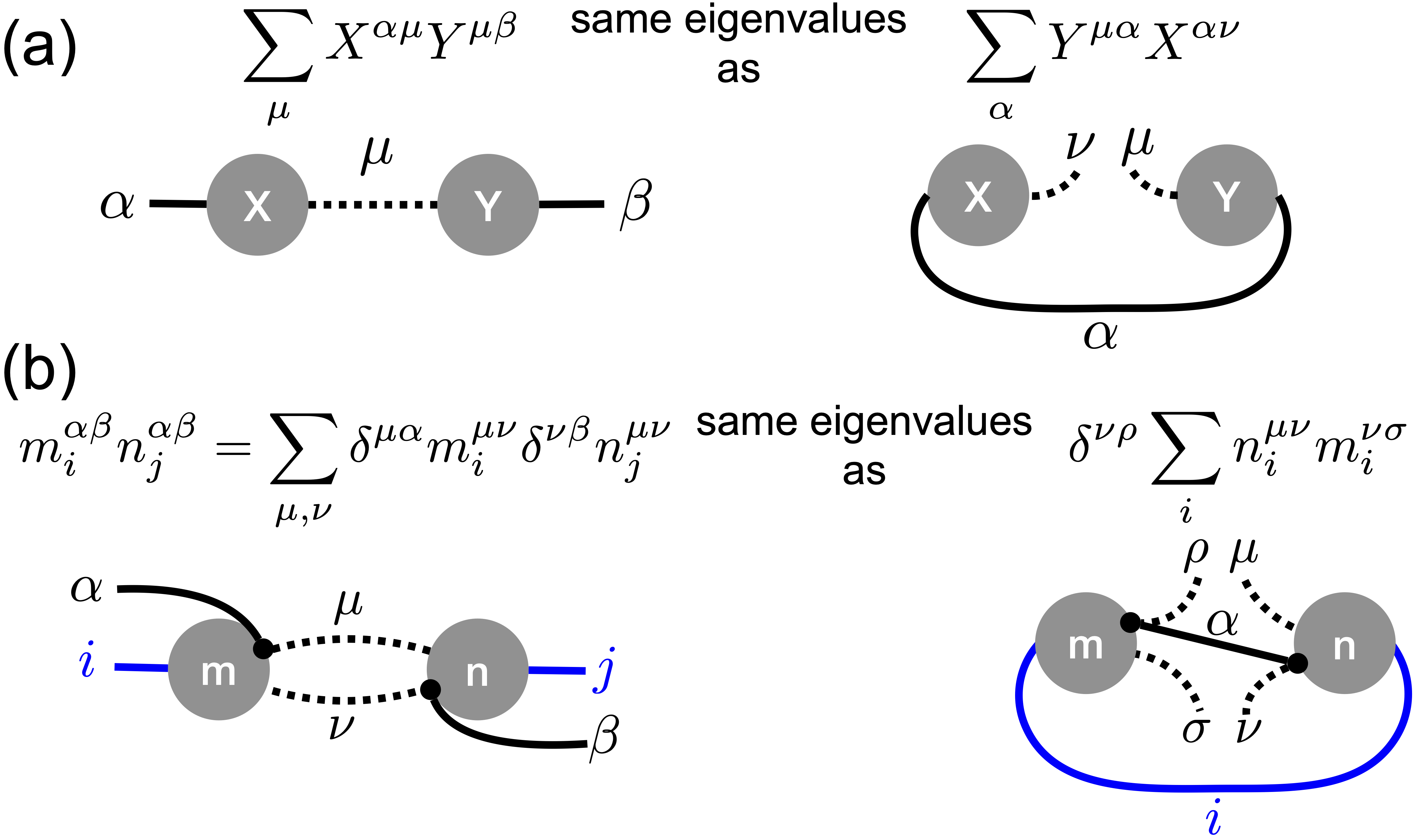}
    \caption{Tensor diagrams \cite{cichocki2014tensor, bridgeman2017hand} illustrating the relationships between (a) matrices $\bm{X}\bm{Y}$ and $\bm{Y}\bm{X}$, and (b) the matrices with more than two indices relevant to $J^{\mu\nu}_{ij}$. In both (a) and (b), the tensors on the left and right have the same eigenvalues, up to zeros, with respect to the left-right bipartition of the indices. Dangling legs are indices of the output tensor, connected legs are summed over, and $\bullet$ is the Kronecker delta. Dashed lines correspond to indices that are summed over on the left and split open on the right. Solid lines are dangling on the left and joined on the right. In (b), black and blue lines sum over $1,\ldots,N$ and $1,\ldots,R$, respectively.}
    \label{fig:tensor}
\end{figure}

We denote the low-rank part of the connectivity by ${L^{\mu\nu}_{ij} = m^{\mu\nu}_i n^{\mu\nu}_j/N}$. This term has up to $R^2$ nonzero eigenvalues, which do not interact with the bulk as $N \rightarrow \infty$; they either are outliers or are swallowed by the bulk. To determine them, we seek an $R^2$-by-$R^2$ matrix whose spectrum coincides with that of $L^{\mu\nu}_{ij}$. Such a matrix can be found using the fact that the matrices $\bm{X}\bm{Y}$ and $\bm{Y}\bm{X}$ have the same spectra up to zeros (Fig.~\ref{fig:tensor}a). We express the low-rank component as $L^{\alpha\beta}_{ij} = N^{-1} \sum_{\mu,\nu} \delta^{\mu \alpha} m_i^{\mu \nu} \delta^{\nu \beta} n_j^{\mu \nu}$,
which contracts over the superindex $(\mu,\nu)$ to form a matrix with superindices $(\alpha,i)$ and $(\beta,j)$. The same eigenvalues, up to zeros, are obtained by contracting over the superindex $(\alpha,i) = (\beta,j)$, resulting in an $R^2$-by-$R^2$ matrix with superindices $(\mu,\nu)$ and $(\rho,\sigma)$,
\begin{equation}
    \hat{T}^{\mu\nu,\rho\sigma} = \delta^{\nu \rho} \frac{1}{N} \sum_i n_i^{\mu\nu} m_i^{\nu\sigma} = \delta^{\nu\rho} T^{\mu \rho \sigma},
    \label{eq:T-hat-def}
\end{equation}
where the limit $N \rightarrow \infty$ was taken in the second step. This can also be derived using a tensor diagram (Fig.~\ref{fig:tensor}b).

When all eigenvalues of $\hat{T}^{\mu\nu,\rho\sigma}$ and the bulk have real parts less than unity, the trivial fixed point of the network is stable, leading to quiescent behavior. If any eigenvalue exceeds this threshold, the network exhibits nontrivial activity described by the DMFT.

\subsection{Analytical evaluation of Gaussian-integral expressions for the error-function nonlinearity}

In the case of the error-function nonlinearity ${\phi(x)= \text{erf}(\sqrt{\pi}x/2)}$, which we use in this paper, the Gaussian integrals in main text Eqs.~11 and~13 can be evaluated analytically to give
\begin{subequations}
\begin{align}
    \psi(\Delta) &= \frac{1}{\sqrt{1 + {\pi \Delta}/{2}}},  \label{eq:psi-erf-expr} \\
   C(\Delta_{12}, \Delta_{11}, \Delta_{22}) &= \frac{2}{\pi}\text{arctan}\left( \frac{\Delta_{12}}{\sqrt{\left(\Delta_{11} + {2}/{\pi}\right)\left(\Delta_{22} + {2}/{\pi}\right) -\Delta_{12}^2}}\right). \label{eq:C-integral-erf}
\end{align}
\end{subequations}

\subsection{Adding cross-region disorder to DMFT equations}

An additional extension to our multiregion model is the inclusion of disorder in the cross-region couplings. The connectivity in this scenario is represented as
\begin{equation}
    J^{\mu\nu}_{ij} = \chi^{\mu\nu}_{ij} + \frac{1}{N} m^{\mu\nu}_i n^{\mu\nu}_j,
\end{equation}
where $\davg{(\chi_{ij}^{\mu\nu})^2 } = G^{\mu\nu}/N$. This combines the model of Aljadeff et al. \cite{aljadeff2015transition} with our communication-subspace model. In this modified system, the current dynamics are unchanged, but the two-point function dynamics (Eq.~9 in main text) and thus the time-dependent gains are updated to
\begin{equation}
    \left(1+\frac{d}{dt}\right)\left(1 + \frac{d}{dt'}\right)\Delta^{\mu}(t, t') = \sum_\nu G^{\mu\nu} C^\mu(t, t') 
    + \sum_{\nu,\rho} U^{\mu\nu\rho}H^{\mu \nu}(t) H^{\mu \rho}(t').
    \label{eq:new-two-pt-fn-dynamics}
\end{equation}
The key new effect is the propagation of high-dimensional fluctuations between regions due to the high-dimensional cross-region connectivity, captured by the coupling of $\Delta^\mu(t,t')$ to other $C^\nu(t,t')$ for $\nu \neq \mu$ in Eq.~\ref{eq:new-two-pt-fn-dynamics}. Thus, the modularity of the disorder-based gating mechanism may not be preserved. Nevertheless, this propagation of fluctuations could shape the current dynamics in interesting ways.

\subsection{Stability analysis via the local energy function}
We start with the local energy given in main text Eq.~20, which involves a first-order perturbation $\sigma^{\mu \nu}\left(t\right)$ around a fixed point $S_0^{\mu \nu}$. We focus on a local energy approach, since we were not able to find a generalized Lyapunov function for the mean-field dynamics with symmetric interactinos.

Computing the time derivative $dE/dt$ and subsequently replacing $\sigma^{\mu\nu} \rightarrow u^\mu \sigma^{\mu\nu}$ gives $dE/dt = \sum_\nu e^\nu$, where 
\begin{equation}
    e^\nu =
    -\sum_\mu (\sigma^{\mu\nu})^2 + b^{\nu}\psi_0^\nu \sum_\mu \sigma^{\nu\mu}\sigma^{\mu\nu}
    + 2b^{\nu}\psi'^\nu_0 \Big{(}\sum_\mu S_0^{\nu\mu}\sigma^{\mu\nu}\Big{)}\Big{(} \sum_\rho S_0^{\nu\rho} \sigma^{\nu\rho}\Big{)}
    +  h^\nu(\sigma^{\nu\nu})^2  \left( \psi_0^\nu + 2\psi'^\nu_0 (S_0^{\nu\nu})^2 \right),
    \label{eq:n-deriv}
\end{equation}
and $\psi^\nu_0 = \psi(A^\nu)$ and $\psi'^\nu_0 = \psi'(A^\nu)$.
We perform a symmetric-antisymmetric decomposition, ${\sigma^{\mu\nu} = \alpha^{\mu\nu} + \beta^{\mu\nu}}$, where ${\alpha^{\mu\nu} = \alpha^{\nu\mu}}$ and ${\beta^{\mu\nu} = -\beta^{\nu\mu}}$, and discard the term $-2\sum_{\mu}\alpha^{\mu\nu}\beta^{\mu\nu}$, which vanishes under the outer $\nu$ sum. 

We then have for a region $\nu$ in \textit{routing mode}
\begin{equation}
    e^\nu =
    2b^\nu \psi'^\nu_0 \Big{(}\sum_\mu S_0^{\mu\nu}\alpha^{\mu\nu} \Big{)}^2
    + h^\nu \psi_0^\nu (\alpha^{\nu\nu})^2
    - 2\sum_\mu (\beta^{\mu\nu})^2 - 2b^\nu \psi'^\nu_0 \Big{(}\sum_\mu S_0^{\mu\nu}\beta^{\mu\nu} \Big{)}^2.
\end{equation}
The first term is nonpositive since $\psi'^\nu_0 < 0$. The second is nonpositive when $h^\nu < 0$, i.e., $a^\nu < b^\nu$. The third and fourth terms, involving $\beta^{\mu\nu}$, are net-nonpositive for all $\beta^{\mu\nu}$ if, and only if, ${- {\psi'(\Delta) \Delta}/{\psi(\Delta)} \leq 1}$. This quantity varies between 0 to $1/2$ as $\Delta$ varies from zero to infinity, so this holds. Thus, $dE/dt \leq 0$. 

For a region $\nu$ in \textit{non-routing mode},
\begin{equation}
    e^\nu = -\left(1 - \frac{b^\nu}{a^\nu}\right)\sum_{\mu\neq\nu}(\alpha^{\mu\nu})^2 + {2a^\nu \psi'^\nu_0 (S_0^{\nu\nu})^2 } (\alpha^{\nu\nu})^2
    - \left(1 + \frac{b^\nu}{a^\nu}\right)\sum_{\mu}(\beta^{\mu\nu})^2.
\end{equation}
The second and third terms are nonpositive, and the first is nonpositive for $a^\nu > b^\nu$. Thus, when the routing and non-routing modes are chosen according to whether $a^\mu$ or $b^\mu$ is larger, the resulting state is stable.

Conversely, if there is a routing mode with ${a^\mu > b^\mu}$, we obtain $dE/dt > 0$ by picking $\alpha^{\mu\mu}$ to be nonzero and all other components of $\alpha^{\mu\nu}$ and $\beta^{\mu\nu}$ to be zero. Similarly, if there is a non-routing mode with $a^\mu < b^\mu$, we obtain $dE/dt > 0$ by picking $\alpha^{\mu\nu}$ to be nonzero and orthogonal to $S^{\mu\nu}_0$ when contracted over $\nu$, and everything else zero. These choices of $\sigma^{\mu\nu}$ indicate directions along which perturbations grow away from the fixed point. For a region $\mu$ in routing mode, there are directions in which the local energy neither grows nor shrinks, $dE/dt=0$, obtained by $\alpha^{\mu\nu}$ being nonzero and orthogonal to $S^{\mu\nu}_0$ when contracted over $\nu$, and everything else zero. We show in the main text that such directions correspond to translation along a continuous attractor manifold.

While $dE/dt<0$ rigorously indicates (marginal) stability, $dE/dt>0$ does not necessarily indicate instability; it might reflect transient dynamics en route to a stable state. Nevertheless, we find through numerical diagonalization of the Jacobian that the perturbations with $dE/dt>0$ given above indeed represent unstable directions. An interesting, as yet unanswered question is whether this system, under the symmetry constraint, possesses a \textit{global} energy function that ensures convergence to fixed points from any initial condition, similar to regular neural networks with coupling symmetry.

\subsection{Dimension and topology of the attractor manifold in multiregion networks with symmetric interactions}

\begin{figure}
    \centering
    \includegraphics[width=3.5in]{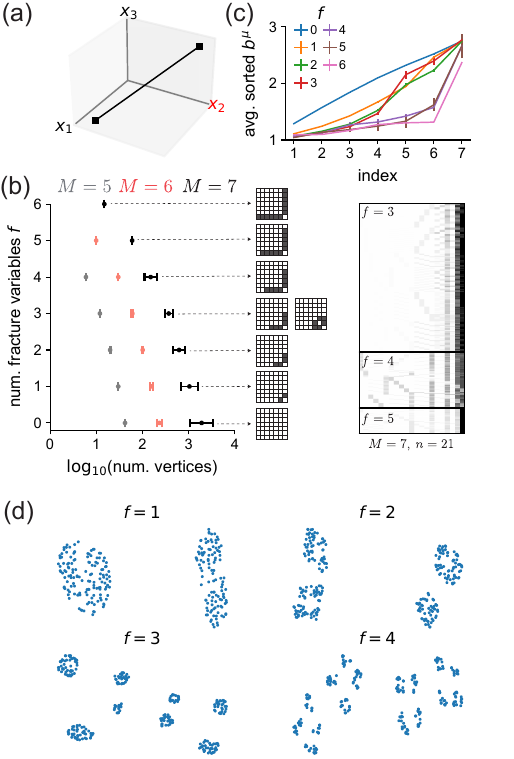}
    \caption{Convex geometry of the attractor manifold in multiregion networks. (a) Cartoon illustration of fracture variables. In this cartoon ``feasible region" defined by the black line segment, $x_2$ is never zero, leading to a binary fracturing of the current-space manifold. (b) Left: Summary of the geometry of the solution polytope, determined by 10000 random choices of $\bm{b}$ (uniform components over [1, 3]). For each random choice, we calculated the number of fracture variables, $f$, and the log-number of vertices. We show the mean and standard deviation of log-number of vertices for each different number of fracture variables. Different colors indicate different numbers of regions in routing mode, $M$. Error bars show 2 standard deviations. Center: Configurations of the current submatrix at fixed points for choices of $\bm{b}$ resulting in $f$ fracture variables, with increasing values of $f$ (ascending) and $M = 7$. Right: ``Barcode'' visualization of all vertices of the feasible region, in $\bm{x}$ space, for specific choices of $\bm{b}$. Each horizontal row corresponds to a different choice of $\bm{b}$, each vertical column corresponds to the $k$-th entry in the vector of variables $\bm{x}$, where $k \in \{1, \ldots, n\}$. The shading indicates the normalized value of $x_k$ at the vertex. Here, $M = 7$, and thus there are $n = 21$ variables in the linear program. Barcodes for $f \in \{3,4,5\}$ are displayed (descending). (c) Averaged sorted values of $b^\mu$, conditioned on yielding specified numbers of fracture variables $f$. (d) t-SNE visualizations of fixed-point manifolds, in current space, for various numbers of fracture variables $f$ with $M = 7$.}
    \label{fig:geometry}
\end{figure}

By imposing symmetry on the effective-interaction tensor, the multiregion system acts as an \textit{attractor network}, with the currents converging to fixed points. These equilibrium states remain unchanged over a timescale significantly longer than that of individual neurons, becoming infinite as $N \rightarrow \infty$. In neuroscience, attractor dynamics have explained memory mechanisms involving discrete and continuous variables, as well as the integration of continuous variables \cite{hopfield1982neural, nair2023approximate, ben1995theory, kim2017ring, chaudhuri2019intrinsic, burak2009accurate, gardner2022toroidal}. Discrete attractors are useful for tasks requiring the retention and recall of specific information, whereas continuous attractors are useful for tasks involving the tracking or integration of ongoing stimuli or movements.

Our analysis thus far has characterized fixed points of the currents without considering the structure of the manifold on which they reside. We now explore this structure through a connection to convex geometry. We show that the architecture of the multiregion network facilitates a blend of discrete and continuous attractors, useful for tasks that necessitate tracking continuous signals in a context-specific way. Furthermore, the dimension and topology of the manifold can be modified by adjusting the effective interactions rather than by rewiring the network architecture. The current-space manifold is linearly embedded in neuronal space, and thus the neuron-space manifold inherits the dimension and topology of the current-space manifold.

We begin by reducing the problem of determining the structure of the current-space manifold to a linear program. For both the disordered and non-disordered cases, the manifold is shaped by three constraints on the submatrix of $S^{\mu\nu}_0$ restricted to the set of regions $\mathcal{S}_\text{route}$ in routing mode: 1) zero on-diagonals; 2) equality constraints on the squared $L^2$-norms of rows (involving $A^\mu$); and 3) the generalized symmetry property, ${S_0^{\mu\nu} u^\nu = S_0^{\nu\mu} u^\mu}$. We encode these constraints using a vector of variables $\bm{x}$ corresponding to the squared upper-triangular elements of the current submatrix restricted to regions in routing mode, i.e., ${\bm{x} = \{(S^{\mu\nu}_0)^2 \: | \: \mu,\nu \in \mathcal{S}_\text{route}, \: \mu < \nu\}}$.
Thus, the number of variables is ${n = M(M-1)/2}$ where $M = |\mathcal{S}_{\text{route}}|$. Crucially, there is a nonnegativity constraint $x_k \geq 0$ for $k\in\{1,\ldots, n\}$ because $x_k$ represents a squared quantity.
There are $M$ linear equality constraints that can be expressed as $\bm{C} \bm{x} = \bm{A}$. Here, $\bm{A}$ has components $A^\mu$, where $A^\mu$ is the required squared $L^2$-norm of row $\mu$ of the current submatrix, and
$\bm{C}$ is an $M$-by-$n$ constraint matrix. Each element of $\bm{C}$ is set to unity or the ratio $b^{\mu}/b^{\nu}$ if the element corresponds to an upper- or lower-triangular element of the current submatrix, respectively; otherwise the element is set to zero (see ``Concrete example of the linear program construction'' below).

The solution set of this linear program, called the \textit{feasible region}, is a convex polytope. Barring fine tuning, its dimension $d$ is the number of variables minus the number of constraints,
\begin{equation}
d = n - M = \frac{M(M-3)}{2}.
\end{equation}
This is also the dimension of the current-space manifold. For $M \geq 4$, the manifold is therefore continuous with dimension $d \geq 2$. For $M = 3$, the manifold reduces to a zero-dimensional point set. For $M = 2$, or for a sufficiently nonuniform constraint vector $\bm{A}$, the linear program is infeasible, i.e., has no solutions. In this case, the assumptions of the linear-program formulation are violated, and the system converges to an exceptional fixed point that can be described analytically by returning to the fixed-point equations (see ``Characterization of the exceptional fixed point'' below).

To characterize the topology of the current-space manifold in non-exceptional cases, we observe that, for a given point $\bm{x}$ on the feasible region, there are $2^n$ corresponding points in current space. This multiplicity arises from the different ways one can choose the signs of the currents. The connectedness, or lack thereof, of the manifold hinges on whether $x_k = 0$ for each $k \in \{1, \ldots, n\}$ is included in the feasible region. If included, positive and negative current-space branches connect; otherwise, a binary fracture of the manifold is induced. We refer to variables $x_k$ that never take on zero as \textit{fracture variables} and denote their number by $f$. This is visualized in Fig.~\ref{fig:geometry}a. Each fracture variable contributes one binary split to the manifold, resulting in $2^f$ connected components. Zeros in components of $\bm{x}$ occur only at vertices of the feasible region, so to identify all fracture variables, it suffices to enumerate all vertices. 


To generate realizations of this linear program, we first pick a vector $\bm{b}$ by sampling its components $b^\mu$ uniformly over $1 \leq b^\mu \leq 3$ and sort them in ascending order for visualization. We set $A^\mu = \psi^{-1}(1/b^\mu)$ for each $\mu$, assuming the disorder-free case with all regions in routing mode. Using the double-description method of Motzkin \cite{motzkin1953double, fukuda1997cdd}, we identify all vertices. We plot $f$ against the log-number of vertices for realizations of the linear program with $M\in\{5,6,7\}$ (Fig.~\ref{fig:geometry}b, left), finding that the number of vertices grows exponentially with $M$ (Fig.~\ref{fig:geometry}b, left). $f$ is negatively correlated with vertex count and is at most $M-1$. Except for $f = 3$, fracture variables correspond to currents for the region with largest $b^\mu$ (Fig.~\ref{fig:geometry}b, center). For $f = 3$, there is an additional configuration involving all currents between the three regions with the largest values of $b^\mu$. We visualized all $n = 21$ vertices for example realizations with $M = 7$ and $f \in \{3,4,5\}$ (Fig.~\ref{fig:geometry}b, right). Choices of $\bm{b}$ leading to more fracture variables tend to have nonuniform components (Fig.~\ref{fig:geometry}c).

We confirmed that the topology predicted by fracture variables matches that of the current-space manifold. For many samples of $\bm{b}$, we evolved the disorder-free DMFT equations from different initial conditions until convergence to fixed points. We then applied t-SNE nonlinear dimensionality reduction to the collection of fixed points \cite{van2008visualizing}. The number of distinct clusters was $2^f$ in all cases, as visualized in Fig.~\ref{fig:geometry}d. Each cluster has some spread corresponding to the continuous dimensions of variation on the manifold.


The dimension and topology of the current-space manifold are determined by the number of regions in routing mode and the values of $A^\mu$, respectively. These quantities can be changed by adjusting $a^\mu$, $b^\mu$, and $g^\mu$. Doing this adjustment dynamically, e.g., through neuromodulation, provides a way to maintain an attractor manifold with a variable dimension and number of connected components, without the need to construct a completely new network architecture. This adaptability could be advantageous for responding to nonstationary tasks or environmental conditions where the computational demands on the attractor system change rapidly and significantly.

\subsection{Concrete example of the linear program construction}
To analyze the dimension and topology of attractors in the case of symmetric effective interactions, we solve a linear program of the form $\bm{C} \bm{x} = \bm{A}$. For concreteness, in the case of $M = 5$, where $M$ is the number of regions in routing mode, symmetric pairs of elements of the current submatrix can be indexed from $1$ through $n = 10$ as
\begin{equation}
\begin{pmatrix}
\bullet & 1 & 2 & 3 & 4 \\
1 & \bullet & 5 & 6 & 7 \\
2 & 5 & \bullet & 8 & 9 \\
3 & 6 & 8 & \bullet & 10 \\
4 & 7 & 9 & 10 & \bullet \\
\end{pmatrix}.
\end{equation}
That is, $x_1 = (S_0^{12})^2 = (S_0^{21})^2$, $x_2 = (S_0^{13})^2 = (S_0^{31})^2$, and so on. This gives the constraint matrix
\begin{equation}
\bm{C} = \begin{pmatrix}
1 & 1 & 1 & 1 & 0 & 0 & 0 & 0 & 0 & 0 \\
\frac{b_2}{b_1} & 0 & 0 & 0 & 1 & 1 & 1 & 0 & 0 & 0 \\
0 & \frac{b_3}{b_1} & 0 & 0 & \frac{b_3}{b_2} & 0 & 0 & 1 & 1 & 0 \\
0 & 0 & \frac{b_4}{b_1} & 0 & 0 & \frac{b_4}{b_2} & 0 & \frac{b_4}{b_3} & 0 & 1 \\
0 & 0 & 0 & \frac{b_5}{b_1} & 0 & 0 & \frac{b_5}{b_2} & 0 & \frac{b_5}{b_3} & \frac{b_5}{b_4} \\
\end{pmatrix}.
\end{equation}

\subsection{Characterization of the exceptional fixed point}
If ${M=2}$, or if the values of $b^\mu$ are highly nonuniform, the linear program is infeasible. Because the trivial fixed point is unstable, there must be at least one stable, nontrivial fixed point in this case that violates the form assumed to parameterize the linear program. We find that this exceptional fixed point is unique up to sign flips and has all current submatrix elements set to zero except for the incoming and outgoing currents in the region with the highest value of $b^\mu$. If this maximal value is $b^M$, this exceptional fixed point can be found by first finding $\psi^M_0$ by solving
\begin{equation}
 \psi^M_0 = \psi\left( \sum_{\mu=1}^{M-1} \frac{\psi^{-1}\left(\frac{1}{b^\mu b^M \psi_0^M}\right)}{b^\mu b^M  (\psi_0^M)^2} \right),
\end{equation}
from which $S_0^{\mu M}$ and $S_0^{M \mu}$ follow. We find that this fixed point becomes stable when the linear program becomes infeasible.

\subsection{Dynamical mean-field theory (DMFT) equations for symmetric effective interactions with disorder}

Assuming stationarity as described in the main text, the DMFT equations become
\begin{subequations}
\begin{align}
    S_0^{\mu\nu} &= \psi^\nu_0 (u^\mu u^\nu + \delta^{\mu\nu}h^\nu) S_0^{\nu\mu}, \\
    \psi_0^\nu &= \psi(\Delta^\nu(0)),\\
    \frac{d^2 \Delta^{\mu}(\tau)}{d\tau^2} &= -\partial_{\Delta^{\mu}} V^\mu(\Delta^{\mu} ; A^\mu), \label{eq:newton} \\
     V^\mu(\Delta^{\mu}; A^\mu) &= -\frac{(\Delta^\mu)^2}{2} 
     + (g^\mu)^2 \Phi(\Delta^\mu, \Delta^\mu(0)) + A^\mu \Delta^{\mu}, \label{eq:potential}\\
    A^\mu&= \sum_\nu (S_0^{\mu\nu})^2,
\end{align}\end{subequations}
where we replaced ${(1 + d/dt)(1 + d/dt') \rightarrow 1 - d^2/d\tau^2}$ in main text Eq.~9, then integrated its rhs with respect to $\Delta^\mu(\tau)$. To obtain $\Phi(\Delta_{12}, \Delta_{11})$, we first define $C(\Delta_{12}, \Delta_{11})$ by setting $\Delta_{11} = \Delta_{12}$ in main text Eq.~10 (see also SI Eq.~\ref{eq:C-integral-erf}), then integrate with respect to $\Delta_{12}$.

These equations generalize main text Eq.~15 to the case where the squared $L^2$-norms of the rows of the current matrix differ from the equal-time two-point functions, ${A^\mu \neq \Delta^\mu(0)}$, due to chaotic fluctuations contributing to the variance of activity in addition to the currents. As in Sompolinsky et al. \cite{sompolinsky1988chaos}, $\Delta^\mu(\tau)$ acts like a Newtonian particle in a Mexican-hat potential, $V^\mu(\Delta^\mu; A^\mu)$. The values of $\psi^\mu_0$ and thus $\Delta^\mu(0)$ are determined by $a^\mu$ and $b^\mu$ as in the disorder-free case, but $A^\mu$ is as yet undetermined. We exchange the dependence of the potential on $A^\mu$ for a dependence on the large-$\tau$ value of the two-point function, ${\Delta^\mu(\infty) = \lim_{\tau \rightarrow \infty} \Delta^\mu(\tau)}$, which satisfies $\left.(dV^\mu/d\Delta^\mu)\right|_{\Delta^\mu(\infty)} = 0$ because the Newtonian particle must come to rest at the top of a hill to obtain a valid decaying two-point function. This condition can be expressed as
\begin{equation}
    A^\mu = \Delta^\mu(\infty) - (g^\mu)^2 C(\Delta^\mu(\infty), \Delta^\mu(0)).
    \label{eq:A_cond}
\end{equation}
We use this to express the potential as $V^\mu(\Delta^\mu; \Delta^\mu(\infty))$, eliminating the dependence on $A^\mu$. Finally, $\Delta^\mu(\infty)$ is determined by energy conservation, $V^\mu(\Delta^\mu(0); \Delta^\mu(\infty)) = V^\mu(\Delta^\mu(\infty); \Delta^\mu(\infty))$. $A^\mu$ can then be found using SI Eq.~\ref{eq:A_cond}, shown in main text Fig.~5(a), and the full form of $\Delta^\mu(\tau)$ is given by integrating the Newtonian dynamics, shown in main text Fig.~5(b). A similar analysis was done by Mastrogiuseppe and Ostojic \cite{mastrogiuseppe2018linking}.


\subsection{Further analysis of asymmetric effective interactions}

\begin{figure}[t]
    \centering
    \includegraphics[width=3.25in]{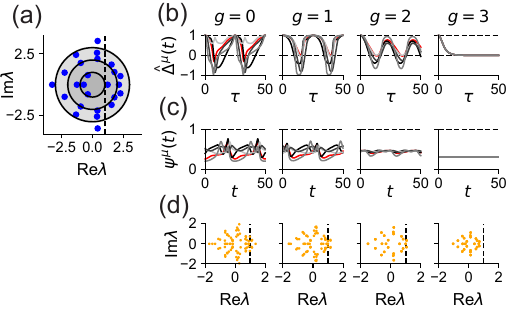}
    \caption{Relationship between disorder and current-variable dynamic complexity in a network of $R = 5$ regions. (a) Spectrum of $\hat{T}^{\mu\nu,\rho\sigma}$. Shaded circles represent the support of the bulk of the spectrum of $J^{\mu\nu}_{ij}$ for disorder levels $g \in \{0,1,2,3\}$. (b--d) Same information is Figs.~\ref{fig:asymm}(b--d), namely, (b) normalized two-point functions, (c) time-dependent gains, and  (d) gain-modulated effective-interaction spectra.}
    \label{fig:fivepop}
\end{figure}

When we allow for fully unconstrained effective interactions, even for the modest values of $R$ considered so far, the dynamics become rich and highly dependent on the specific form of the effective-interaction tensor. This raises the question of whether we can glean general insights when the number of regions is large, as is the case in real neural circuits. To investigate this, we examine a model of $R=5$ regions and asymmetric effective interactions. We randomly sample an effective-interaction tensor that yields complex-conjugate leading eigenvalues, along with several other real and complex unstable modes (Fig.~\ref{fig:fivepop}a). Without disorder, this effective-interaction tensor produces an intricate limit cycle in the currents (Fig.~\ref{fig:fivepop}b, $g=0$). Increasing the disorder variance parameter $g$ uniformly across regions reduces the complexity of the limit cycle due to high-dimensional fluctuations disrupting communication between regions (Fig.~\ref{fig:fivepop}b, $g=1$ and $g=2$). For sufficiently large $g$, the currents vanish as disordered connectivity within regions overtakes structured communication (Fig.~\ref{fig:fivepop}b, $g=3$). The gradual transition from a complex to a simple limit cycle, and eventually to its absence with increasing disorder, can be understood in terms of the growing radius of the spectral bulk, which swallows more and more outlier modes linked to the dynamics of the currents (Fig.~\ref{fig:fivepop}a, circles). Inspection of the gains $\psi^\mu(t)$ reveals that, rather than vanishing, gains remain of order unity (Fig.~\ref{fig:fivepop}c). However, gains exhibit less complexity across time for larger $g$. Moreover, the leading eigenvalues of the spectrum of $\psi^\nu(t) \hat{T}^{\mu\nu,\rho\sigma}$ hover around unity, with a diminishing number of modes crossing the stability line as $g$ increases (Fig.~\ref{fig:fivepop}d).

\subsection{DMFT with inputs}

We extend our analysis to include scalar inputs $I^\mu(t)$ for each region $\mu$. These inputs are supplied along a vector $\bm{v}^\mu$, whose components are jointly Gaussian with all other connectivity vectors in the system. The high-dimensional dynamics now become:
\begin{equation}
    \frac{dx^\mu_i(t)}{dt} = -x^\mu_i(t) + \sum_{\nu=1}^R \sum_{j=1}^N J^{\mu\nu}_{ij} \phi^\nu_j(t) + v^\mu_i I^\mu(t)
\end{equation}
Applying the DMFT analysis to this extended system yields the following expanded dynamics for the currents and two-point function:
\begin{align}
    \frac{dS^{\mu\nu}(t)}{dt} &= -S^{\mu\nu}(t) + \psi^\nu\left(t\right) \sum_{\rho=1}^R T^{\mu\nu\rho} S^{\nu\rho}(t) + V^{\mu\nu} I^\nu(t), \\
    \left(1 +\frac{d}{dt}\right)\left(1 +\frac{d}{dt'}\right)\Delta^\mu(t, t') &= (g^\mu)^2 C^\mu(t, t') + \sum_{\nu,\rho=1}^R U^{\mu\nu\rho} H^{\mu\nu}(t) H^{\mu\rho}(t') \nonumber \\
    &+ w^\mu I^\mu(t) I^\mu(t') + \sum_{\nu=1}^R W^{\mu\nu} \left[  H^{\mu\nu}(t) I^\mu(t') + H^{\mu\nu}(t') I^\mu(t) \right].
\end{align}
Here, we introduce new parameters $V^\mu$, $w^\mu$, and $W^{\mu\nu}$, defined as averages:
\begin{align}
    V^{\mu\nu} &= \davg{n^{\mu\nu}_i v^\nu_i}, \\
    w^\mu &= \davg{\left(v^\mu_i\right)^2}, \\
    W^{\mu\nu} &= \davg{m^{\mu\nu}_i v^\mu_i}.
\end{align}

\begin{figure}[t]
    \centering
    \includegraphics[width=3.25in]{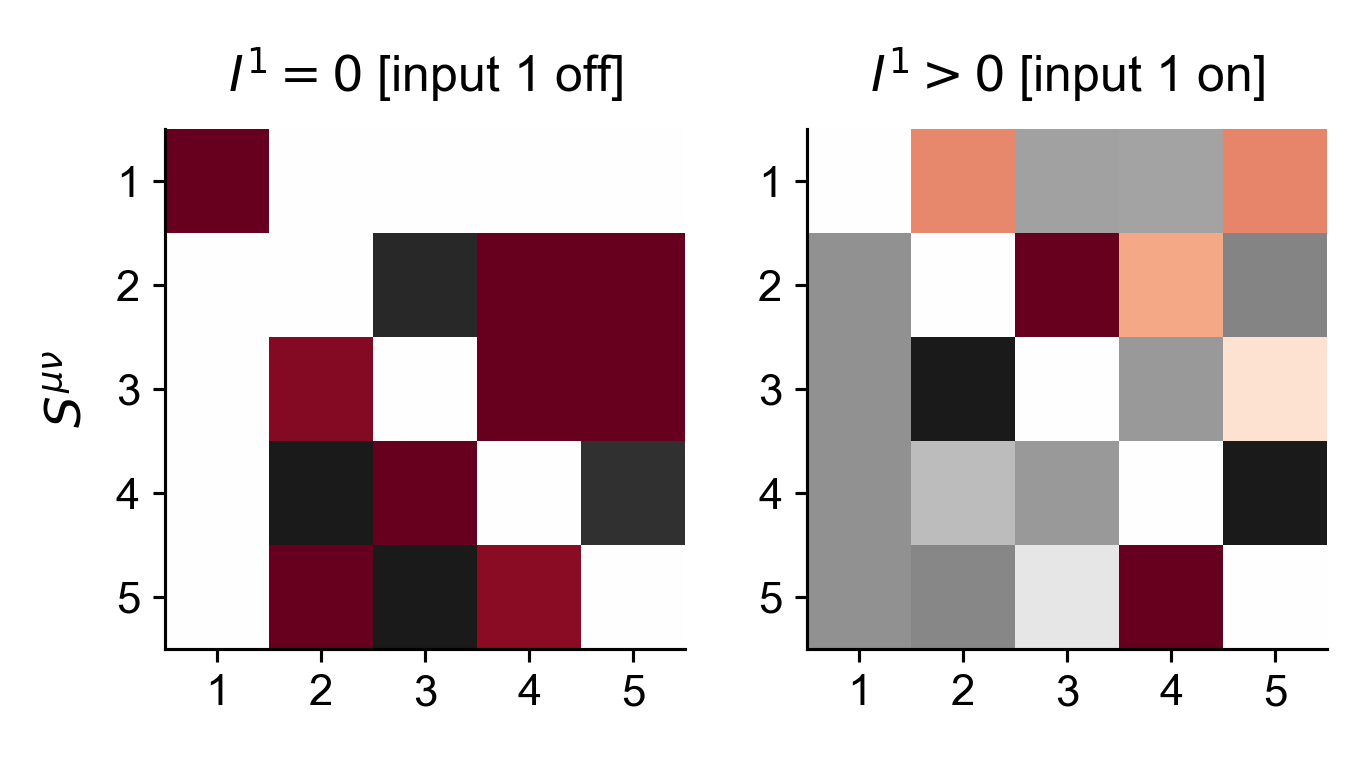}
    \caption{Input-driven switching in a 5-region network. Left: No input ($I^1 = 0$). Right: Strong input ($I^1 = 2$). The transition shows region 1 changing from self-excitation without routing to routing without self-excitation.}
    \label{fig:input-flip}
\end{figure}

To demonstrate input-driven switching, we consider a simple scenario with $R=5$ regions, using the same parameterization of connectivity as in the main text Fig. 3 (middle). We activate an input in region 1, while all other inputs remain zero. The parameters are set as follows:
\begin{equation}
    V^{\mu\nu} = 0.1 \times \begin{pmatrix}
0 & 0 & 0 & 0 & 0 \\
1 & 0 & 0 & 0 & 0 \\
1 & 0 & 0 & 0 & 0 \\
1 & 0 & 0 & 0 & 0 \\
1 & 0 & 0 & 0 & 0
\end{pmatrix}, \quad w^\mu = 1, \quad W^{\mu\nu} = 0.
\end{equation}
Fig.~\ref{fig:input-flip} illustrates the effect of varying the input strength in region 1 from $I^1 = 0$ (left) to $I^1 = 2$ (right). We observe a transition in which region 1 switches from exciting itself without routing to not exciting itself and routing. We note that there is an intermediate region, appearing immediately when $I^1$ becomes nonzero, where the input and output currents from region 1 become nonzero, but the self-excitation of region 1 persists.
This example demonstrates how external inputs can modulate the routing behavior in multiregion networks, providing a mechanism for flexible, context-dependent information processing.

\subsection{Extension of multiregion networks to higher-rank communication subspaces with mixture-of-Gaussians loadings.}
\begin{figure}[b]
    \centering
    \includegraphics[width=3.5in]{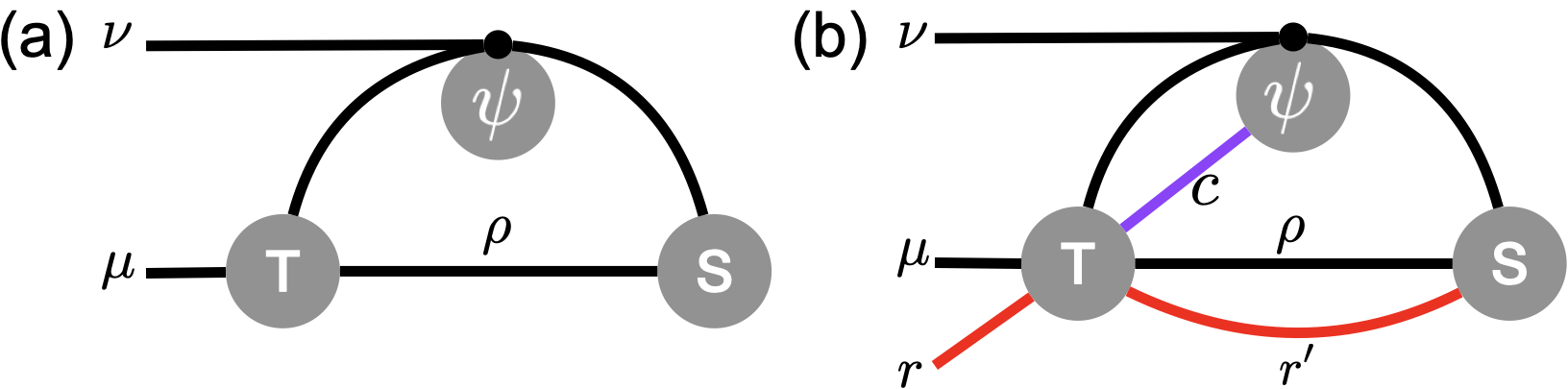}
    \caption{Tensor diagrams illustrating the extension of multiregion networks to higher-rank communication subspaces with mixture-of-Gaussians loadings. (a) Form of $(1 + d/dt)S^{\mu\nu}(t)$ in the multiregion network we have studied. (b) Form of $(1+d/dt)S^{\mu\nu}_r(t)$ in the aforementioned extension (in this case, the currents acquire an additional $r$ index that runs over rank-one components). The diagram has two new internal lines corresponding to contractions over mixture components ($c$) and rank-one components ($r'$).}
    \label{fig:tensor-diagram-2}
\end{figure}

See Fig.~\ref{fig:tensor-diagram-2}.

\subsection{Locating multiregion networks in the space of low-rank mixture-of-Gaussians networks}

Setting aside disorder, our model involves a blockwise low-rank coupling matrix, an embodiment of a broader idea where neurons are assigned group identities and coupling statistics are based on these identities. Another embodiment of this idea is the low-rank mixture-of-Gaussians model proposed by \cite{beiran2021shaping, dubreuil2022role}, where the coupling matrix is a sum of rank-one outer products with mixture-of-Gaussians loadings. In this framework, each neuronal group corresponds to a Gaussian mixture component. Our multiregion network model, of $R$ regions, is a special case of a low-rank mixture-of-Gaussians network with rank $R^2$ and $R$ mixture components.

Here, we demonstrate via an explicit construction that multiregion networks are a special case of the low-rank mixture-of-Gaussians model.
Consider a low-rank network where the rank-one terms are indexed by $r$ (or $r'$), neurons are indexed by $i$ (or $j$), and the coupling matrix $W_{ij}$ is defined as
\begin{equation}
W_{ij} = \sum_r v^r_i w^r_j.
\end{equation}
The components of the vectors $v^r_i$ and $w^r_i$ follow a mixture-of-Gaussians distribution with i.i.d. sampling across the neuron index, $i$. Each mixture component has zero mean. The second-order statistics are defined by
\begin{align}
\tavg{w^r_i v^{r'}_i}_c &= C  t^{rr'}[c], \\
\tavg{v^r_i v^{r'}_i}_c &= C  u^{rr'}[c],
\end{align}
where $\tavg{\cdot}_c$ denotes an average in mixture component $c$ and $C$ is the number of mixture components. We assume that all mixture components have equal probability. With these definitions, the mean-field equations were shown in \cite{beiran2021shaping, dubreuil2022role} to be
\begin{align}
\left(1 + \frac{d}{dt}\right)\kappa^r(t) &= \sum_{r'} \left[ \sum_c \psi^c(t) t^{r r'}[c] \right] \kappa^{r'}(t), \\
\psi^c(t) &= \psi\left(\sum_{r,r'} u^{rr'}[c] \kappa^r(t) \kappa^{r'}(t)\right),
\end{align}
where $\psi(\Delta)$ is given for the error-function nonlinearity by SI Eq.~\ref{eq:psi-erf-expr}. 

Toward making a multiregion network emerge from these equations, we consider $R^2$ rank-one terms and $R$ mixture components, and substitute $r \rightarrow (\mu,\nu)$, $r' \rightarrow (\rho,\sigma)$, and $c \rightarrow \alpha$ (the purpose of the last replacement is simply to use a Greek letter for consistency). The second-order statistics are constructed as follows:
\begin{align}
t^{\mu\nu,\rho\sigma}[\alpha] &= \delta^{\alpha \nu} \delta^{\alpha \rho} T^{\mu\nu\sigma}, \\
u^{\mu\nu,\rho\sigma}[\alpha] &= \delta^{\alpha \mu} \delta^{\alpha \rho} U^{\alpha \nu \sigma},
\end{align}
where $T^{\mu\nu\rho}$ and $U^{\mu\nu\rho}$ are the tensors defining the multiregion network of interest. Under this construction, the mean-field equations transform into those for multiregion networks.

We have reproduced the mean-field equations of the multiregion network, but do realizations of the couplings $W_{ij}$ exhibit the blockwise low-rank structure of multiregion networks? Consider the rank-one term $v^{\mu\nu}_i w^{\mu\nu}_j$ in this construction. Note that $\tavg{(v^{\mu\nu}_i)^2}_\alpha$ is proportional to $\delta^{\alpha \mu}$. Thus, only the rows corresponding to mixture component $\mu$ are nonzero in this rank-one term. Now, the second-order statistics among the ``$w$'' vectors are not relevant to the mean-field equations. But, if we assume that $\tavg{(w^{\mu\nu}_i))^2}_\alpha$ is proportional to $\delta^{\alpha \nu}$, only the columns corresponding to mixture component $\nu$ are nonzero in this rank-one term. Thus, in rank-one term $(\mu,\nu)$, the nonzero entries form a submatrix corresponding to rows in mixture component $\mu$ and columns in mixture component $\nu$. Given that there is a single rank-one term for each $(\mu,\nu)$, a rank-one submatrix is present at every $(\mu,\nu)$ block of $W_{ij}$.

In summary, we have located multiregion networks in the space of low-rank mixture-of-Gaussians networks. In particular, when the $u^{rr'}[c]$ and $U^{\mu\nu\rho}$ terms are fixed, multiregion networks lie on an $R^3$-dimensional manifold in the $R^5$-dimensional space of rank-$R^2$ networks with $R$ mixture components. In this sense, multiregion networks possess a high degree of structure compared to generic networks in the low-rank mixture-of-Gaussians class.

Furthermore, as described in the main text, multiregion networks can themselves be generalized to have rank-$K$ communication subspaces and $C$ Gaussian-mixture components. This extension can be captured by a low-rank mixture-of-Gaussians construction with $KR^2$ rank-one terms and $CR$ mixture components. Compared to generic networks in the low-rank mixture-of-Gaussians class, multiregion networks possess a far greater degree of structure. An interesting question is whether the inductive bias corresponding to multiregion networks is advantageous in constructing models within this class.


\newpage 
\bibliography{refs}

\end{document}